%% file: main.tex
\documentclass[journal]{IEEEtran}

\ifCLASSINFOpdf

\else

\fi

\usepackage{cite}
\usepackage{amsmath,amssymb,amsfonts}
\usepackage[dvipsnames]{xcolor}
\usepackage{graphicx}
\usepackage{textcomp}

\usepackage[english]{babel}
\usepackage{blindtext}
\usepackage[font=footnotesize]{caption}
\usepackage[font=footnotesize]{subcaption}
\usepackage{soulutf8}
\usepackage[english]{babel}
\usepackage[utf8]{inputenc}
\usepackage{colortbl}
\usepackage[linesnumbered,algo2e,ruled]{algorithm2e}
\usepackage[inner=2.5cm,outer=2.5cm,top=3cm,bottom=3cm]{geometry}
\usepackage{tabularx}
\usepackage{algorithm}
\usepackage[thinlines]{easytable}
\usepackage{algpseudocode}
\usepackage{svg}
\usepackage{bm}
\usepackage{array}
\newcolumntype{?}{!{\vrule width 1.5pt}}

\usepackage{tablefootnote}
\usepackage{longtable}
\usepackage{multirow}
\usepackage{booktabs}
\usepackage{soul}
\usepackage[hyphens]{url}
\usepackage[colorlinks=true,linkcolor=blue,breaklinks=True,citecolor=brown,urlcolor=blue]{hyperref}
\usepackage{siunitx}
\usepackage{amsthm}
\usepackage{booktabs,subcaption,dcolumn}

\usepackage{mathtools, nccmath}
\usepackage{pifont}

\usepackage{booktabs,subcaption,amsfonts,dcolumn}
\newcolumntype{d}[1]{D..{#1}}

\newcommand{\POC}{{\small{\textsf{POC}}}}

\newcommand{\LNU}{{{\small{\textsf{LNU-Phish}}}}}
\newcommand{\POCscript}{{\scriptsize{\textsf{POC}}}}
\newcommand{\LNUscript}{{{\scriptsize{\textsf{LNU-Phish}}}}}

\newcommand{\res}[1]{{\footnotesize{$#1$}}}

\newcommand{\cmark}{\color{ForestGreen}{\ding{51}}}%
\newcommand{\xmark}{\color{RubineRed}{\ding{55}}}%
\newcommand{\nop}[1]{{}}
\newcommand{\GBA}[1]{{\textsf{GBA-#1}}}
\def\Fmop{{\mathsf{Fmop}}}

\newcommand{\overbar}[1]{\mkern 1.5mu\overline{\mkern-1.5mu#1\mkern-1.5mu}\mkern 1.5mu}

\begin{document}

\title{Mitigating Adversarial Gray-Box Attacks Against Phishing Detectors}

\author{
    \IEEEauthorblockN{Giovanni Apruzzese\IEEEauthorrefmark{1}, V. S. Subrahmanian\IEEEauthorrefmark{2}}\\
    
    \IEEEauthorblockA{\IEEEauthorrefmark{1}{\small \textit{University of Liechtenstein}, Hilti Chair of Data and Application Security---Vaduz, Liechtenstein}\\
    {\small giovanni.apruzzese@uni.li}}\\
    
    \IEEEauthorblockA{{\small \IEEEauthorrefmark{2}\textit{Northwestern University}, Dept. of Computer Science \& Buffett Institute of Global Affairs---Evanston, IL, USA}\\
    {\small vss@northwestern.edu}}
}

\markboth{IEEE Transactions on Dependable and Secure Computing}%
{Shell \MakeLowercase{\textit{et al.}}: Bare Demo of IEEEtran.cls for IEEE Journals}

\maketitle

\input{sections/0-abstract}

\begin{IEEEkeywords}
phishing detection, cybersecurity, adversarial attacks, websites, dataset
\end{IEEEkeywords}

\IEEEpeerreviewmaketitle

\input{sections/1-introduction.tex}
\input{sections/2-related.tex}

\input{sections/3-LNU_dataset.tex}

\input{sections/4-attacks.tex}
\input{sections/5-defense.tex}
\input{sections/6-experiments.tex}
\input{sections/7-discussion.tex}
\input{sections/8-prevalence.tex}
\input{sections/9-conclusions.tex}

\ifCLASSOPTIONcaptionsoff
  \newpage
\fi

\input{main.bbl}


\input{biographies.tex}

\end{document}

%% file: sections/0-abstract.tex
\begin{abstract}
Although machine learning based algorithms have been extensively used for detecting phishing websites, there has been relatively little work on how adversaries may attack such ``phishing detectors'' (PDs for short). In this paper, we propose a set of Gray-Box attacks on PDs that an adversary may use which vary depending on the knowledge that he has about the PD. We show that these attacks severely degrade the effectiveness of several existing PDs. We then propose the concept of \emph{operation chains} that iteratively map an original set of features to a new set of features and develop the ``Protective Operation Chain'' (\POC\ for short) algorithm. \POC\ leverages the combination of random feature selection and feature mappings in order to increase the attacker's uncertainty about the target PD. Using 3 existing publicly available datasets plus a fourth that we have created and will release upon the publication of this paper\footnote{We provide a sample of our dataset for the referees. We release our resources at: \url{https://lnu-phish.github.io/}}, we show that \POC\ is more robust to these attacks than past competing work, while preserving predictive performance when no adversarial attacks are present. Moreover, \POC\ is robust to attacks on 13 different classifiers, not just one. These results are shown to be statistically significant at the $p < 0.001$ level.
\end{abstract}

%% file: sections/1-introduction.tex

\section{Introduction}
\label{sec:introduction}

\IEEEPARstart{M}{achine} learning algorithms are increasingly used in a wide array of cybersecurity applications including malware detection~\cite{maiorca2018towards}, intrusion detection~\cite{buczak2016survey}, insider threat detection~\cite{mayhew2015use}, spam detection~\cite{zhao2017efficient}, and the detection of phishing websites~\cite{subasi2017intelligent}.

Phishing attacks are one of the most common types of attacks. ProofPoint's 2020 ``State of the Phish'' report\footnote{\url{https://proofpoint.com/us/resources/threat-reports/state-of-phish}}
states that over $1.5$ million phishing websites are created \emph{every month} and that $90$\% of businesses reported being a victim of a phishing attack in 2019. Phishing attacks offer one of the easiest ways for malicious hackers to penetrate an enterprise. Considerable work has gone into addressing this problem~\cite{moore2011impact,basnet2014learning, abdelhamid2014phishing,abdelhamid2017phishing,verma2015character,jeeva2016intelligent,tan2016phishwho,subasi2017intelligent,niakanlahiji2018phishmon,ali2017phishing,lancaster2018maltp}. In addition to blacklists maintained by corporations such as Google, there are also publicly available blacklists from sites such as PhishTank\footnote{\url{https://www.phishtank.org/}}. 
However, these sources become obsolete frequently as malicious hackers move their phishing URLs from site to site in order to evade detection. Rule-based systems were therefore developed by several researchers. For instance,~\cite{cook2008phishwish} develops a set of $8$ rules to capture phishing webpages while other approaches analyze the content of a website~\cite{zhang2007cantina}. Another effort~\cite{chen2009fighting} has examined the use of a discriminative set of features associated with phishing URLs and then checked to see whether a given URL was similar to a known phishing URL based on their associated feature vectors. \cite{medvet2008visual,hara2009visual} pioneered the idea of using visual similarity between a legitimate web page (e.g. a bank website) and another website in order to check if the latter might be a phishing website. During the past decade, using machine learning to detect phishing has become widespread. Early efforts in this direction include~\cite{kim2011detecting, liu2010automatic,zhang2011textual,whittaker2010large,cranor2006evaluation,bergholz2010new,toolan2009phishing}. Despite even more recent research efforts proposing increasingly sophisticated machine learning solutions to counter this threat~\cite{moore2011impact,basnet2014learning, abdelhamid2014phishing,abdelhamid2017phishing,verma2015character,jeeva2016intelligent,tan2016phishwho,subasi2017intelligent,niakanlahiji2018phishmon,ali2017phishing}, phishing websites still represent a dangerous menace~\cite{kettani2019threats} as is evident from the aforementioned ProofPoint report.
 
One reason for this is that machine learning classifiers are trained on a training dataset from which they learn a model that separates benign entries from illegitimate ones. However, adversaries (i.e. malicious hackers) are continuously adapting to Phishing Detectors (PDs for short). Often times, these adaptations are very simple, allowing their phishing webpages to evade existing PDs with relative ease. Most work on adversarial machine learning for cyber-security deals with two extremes: ``white box'' attacks in which the adversary has full knowledge of the defenses used by the PD (e.g. classifier used, list of features used) or ``black box'' attacks in which the adversary has no knowledge whatsoever. Real world attackers are not likely to have knowledge that is at either of these extremes --- their knowledge is most likely somewhere in the ``gray area'' between full knowledge (white box) and no knowledge (black box) of the defenses. We use the term ``Gray Box'' to refer to attacks where the attacker's knowledge can lie between these two extremes. 
 
We propose a more robust phishing website detector that is capable of withstanding a large class of Gray Box attacks.\footnote{Please note that we do NOT claim robustness to all types of Gray Box attacks, but only to certain classes defined in the paper.} Our notion of Gray Box is powerful enough to capture both White Box and Black Box settings. In particular, we make the following contributions.
 
\begin{itemize}
    
    \item \emph{Gray Box Attack Scenarios.}  We consider two types of Gray Box attacks: \textit{simple} attacks, where the adversary only knows and modifies a few features and \textit{complex} attacks, where the adversary knows a percentage $\Delta$ of the set of features used by the defender. When $\Delta=100\%$, we have a White Box attack. When $\Delta=0\%$, we have a Black Box attack.
    
    \item We show that phishing detectors based on \textbf{13} recent works~\cite{mohammad2014predicting,abdelhamid2014phishing,verma2015character,jeeva2016intelligent,abdelhamid2017phishing,subasi2017intelligent,ali2017phishing,corona2017deltaphish,babagoli2019heuristic,jain2018towards,niakanlahiji2018phishmon,sahingoz2019machine, basnet2014learning} during the 2014--2019 timeframe are susceptible to these attacks. Specifically, we formally define the \emph{Impact} of an attack on a phishing detector and show that these attacks induce a statistically significant impact (performance reduction) against these 13 well-known classifiers, including 2 deep learning classifiers.
    Thus, our Gray Box attacks encompass the above 2 attack scenarios and apply to 13 classifiers (as opposed to just one as in most past work).\footnote{Like past work in the area, we do not handle attacks on custom classifiers that are often kept secret.} The impact of these Gray Box attacks is shown on 3 well-known datasets (DeltaPhish~\cite{deltaphish2018dataset}, Mendeley~\cite{mendeley2018dataset} and UCI~\cite{UCI2015dataset}) as well as a fourth (new) dataset that we have created.
    
    \item We define the notion of an \emph{operation chain} ($oc$). Operation chains transform existing samples into a new feature space through the iterative application of some simple operators. Even if the adversary knew the original feature set, he is unlikely to know such new feature space\footnote{Unless of course he has already compromised the enterprise system, but in that case, he would not need to phish employees of the enterprise!}. We propose the \emph{Protective Operation Chain} (\POC) algorithm. 
    
    \item We show that \POC\ is more robust to these attacks than 13 existing PDs on the same 4 datasets mentioned above. Past works use a wide range of classification techniques and features---so \POC\ is robust when used on top of many different classification algorithms and feature sets. To validate the claim that \POC\ outperforms existing baselines, we carry out a very rigorous Wilcoxon Signed Rank test (which imposes tougher metrics than, e.g., simple t-test) along with a Bonferroni correction. Such test demonstrates that \POC\ increases the robustness of past baselines at the $p\!\!<\!\!0.001$ level, i.e. the probability that \POC\ really outperforms past work (as opposed to doing so by accident) is over $99.9\%$.\footnote{We note that $p < 0.05$ is the common standard for a one-star claim of statistical significance, $p < 0.01$ is the standard used for a 2-star claim, and $p<0.001$ is the standard used for a 3 star claim. Our results put \POCscript\ well within this highest statistical significance category.}
    
    \item We show that \POC\ is practical for real deployments. We show that the using \POC\ causes a statistically negligible performance degradation in the \textit{absence} of attacks (with respect to the baselines). We conduct an in-depth analysis showing the pros and cons of \POC\ when used to harden the best baseline PD for each dataset.
    
    \item A final contribution is our new dataset, \LNU\ (short for Liechtenstein and Northwestern University-Phishing). \LNU\ overcomes several problems affecting existing datasets which we discuss shortly and can serve as a ``future-proof'' benchmark for developing novel PDs. We will release both \LNU\ and the code to compute our \POC\ implementation after paper acceptance.
    
\end{itemize}
 
The remainder of this paper is structured as follows. Section~\ref{sec:related} presents related work. Section~\ref{sec:LNU} motivates our \LNU\ dataset and explains how it was built. Section~\ref{sec:attacks} outlines the adversarial Gray Box attacks proposed in this paper. The description of operation chains and the \POC\ algorithm is provided in Section~\ref{sec:defense}.
Section~\ref{sec:experiments} shows our main experimental results, which are formally analyzed and discussed in Section~\ref{sec:discussion}. Section~\ref{sec:prevalence} provides additional experiments on a special application of \POC.
Section~\ref{sec:conclusions} concludes the paper and suggests avenues for future work.

%% file: sections/2-related.tex
\section{Related work}
\label{sec:related}

We divide related work into 3 parts: (i)~detection of phishing websites; (ii)~vulnerability of ML to adversarial attacks and existing countermeasures; (iii)~adversarial attacks against phishing detectors (PD).

\subsection{Detection of Phishing Websites} 
\label{sec:phishing_detection}
Though rule based methods (e.g.~\cite{moore2011impact}) were initial used to detect phishing sites, machine learning (ML) based approaches are now common. ~\cite{basnet2014learning,verma2015character,ma2009beyond,wang2015breaking} identify phishing URLs by analysing hundreds of features extracted from the corresponding URLs. Other approaches~\cite{mohammad2014predicting, abdelhamid2014phishing, verma2015character, jeeva2016intelligent, sahingoz2019machine} develop classifiers that use a reduced number of URL-based features while achieving similar or superior accuracy (e.g., over $97\%$ in~\cite{sahingoz2019machine}). 
~\cite{garera2007framework} suggests using information obtained by external sources (e.g., DNS logs) as features. Some papers combine URL-based features with HTML-based features to improve performance\footnote{Because the word ``accuracy'' has a specific technical meaning in machine learning, we will use the term ``performance'' to refer to the quality of results generated by a classifier.}~\cite{abdelhamid2017phishing,subasi2017intelligent,ali2017phishing,babagoli2019heuristic, tan2016phishwho}. ~\cite{jain2018towards} and~\cite{niakanlahiji2018phishmon} also consider information provided by external reputation sources (such as DNS records). More recently,~\cite{corona2017deltaphish, tian2018needle} leverage image processing with HTML inspection to detect phishing content in compromised websites.~\cite{lancaster2018maltp} develops methods to predict how Twitter is used to lure victims to phishing sites.
Despite all these efforts, phishing websites still represent a widespread menace~\cite{kettani2019threats}. 

\subsection{Adversarial Machine Learning}
\label{sec:adversarial_settings}
The recent success of deep learning has led to work showing that small perturbations to the input can lead to huge errors in image processing~\cite{carlini2017towards, papernot2016limitations, goodfellow2014explaining, chen2017zoo, su2019one} as well as in text and speech processing~\cite{carlini2016hidden, jia2017adversarial}. There has also been important work in general cybersecurity~\cite{biggio2018wild,katzir2018quantifying,kantarcioglu2016adversarial,shi2017evasion,vorobeychik2018adversarial}. 

However, most existing work on adversarial ML makes very strong assumptions about the adversary's knowledge about the defense. Such knowledge is typically denoted with the notion of `box' threat models.
White box models assume the attacker has complete knowledge of the defense including the algorithm used and all the features used~\cite{warzynski2018intrusion, chen2018automated, munoz2017towards, grosse2017adversarial}. Conversely, black box models~\cite{rosenberg2018generic, papernot2017practical, dang2017evading, xu2016automatically, li2014feature} assume the adversary knows absolutely nothing about the target's defenses. Both of these are extreme cases---in the real-world, defenders might use `customized' classifiers (e.g. ensembles) with novel features and feature selection and late fusion~\cite{bai2019dbank} or custom combinations of supervised and unsupervised learning~\cite{chakraborty2017ec2} which would be almost impossible for an adversary to guess correctly.

Some recent work considers other scenarios~\cite{demontis2017yes,apruzzese2018evading,anderson2016deepdga}, but assume the classifier used by the PD is known which is unlikely~\cite{pendlebury2018enabling, gardiner2016security}. 
Some papers~\cite{papernot2016distillation,grosse2017adversarial} propose methods to harden detectors based on Neural Networks, while~\cite{kantchelian2016evasion} proposes an approach to improve the robustness of tree-based mechanisms; other efforts focus on SVM-based techniques~\cite{biggio2015one, russu2016secure, he2017robust}. 

Our \POC\ algorithm improves upon past work in the following ways: (i) we are the first to propose mapping feature vectors into a new feature space for purposes of increasing robustness to adversarial attacks against phishing detectors\footnote{Note that mapping feature spaces into new feature spaces is not new in other domains --- for example, kernel tricks used in Support Vector Machines leverage a similar strategy.}, (ii) \POC\ is experimentally shown to be robust against attacks on 13 different classifiers as opposed to just one, (iii) \POC\ is robust to different variants of Gray Box attacks, and (iv) we carry out our evaluation on 4 different datasets---not just 1.

\subsection{Adversarial Attacks Against Phishing Detectors}
\label{sec:attacks_pd}
Most work on adversarial ML in cybersecurity has focused on malware detection~\cite{maiorca2018towards}, spam detection~\cite{zhao2017efficient, russu2016secure, biggio2013security, he2017robust} and network intrusions~\cite{buczak2016survey, apruzzese2021modeling}. An important recent effort~\cite{liang2016cracking} reverse engineers and subverts the phishing detector used by the Google Chrome web browser. Another important paper~\cite{corona2017deltaphish} devises a PD by combining the analysis of the webpage HTML and its image data in a white box setting where the attacker has complete control of the entire webpage domain.

Both of these efforts, though very important, make strong assumptions:~\cite{liang2016cracking} attacks just \emph{one} phishing detector (albeit an important one); whereas~\cite{corona2017deltaphish} only assumes white box adversary that attacks only one classifier (linear SVM). In contrast, \POC\ hardens multiple classifiers used by a defender and can protect against multiple attack models. Furthermore, \POC\ is robust to attacks against 13 different classifiers including recent ones (e.g., Google's Deep \&\ Wide), as opposed to just one classifier considered in most previous works. Finally, as mentioned earlier, \POC's performance is tested on 4 datasets, not just one, using distinct (but similar) feature sets.

%% file: sections/3-LNU_dataset.tex
\section{The \LNU\ Dataset}
\label{sec:LNU}

There are a number of well-known existing datasets for phishing website detection. They include
Mendeley~\cite{mendeley2018dataset}, DeltaPhish~\cite{deltaphish2018dataset},  UCI~\cite{UCI2015dataset},
PhishStorm~\cite{marchal2014phishstorm} and Ebbu\footnote{\url{https://github.com/ebubekirbbr/pdd}}.

\subsection{Problems with Existing Phishing Datasets.} 
Most existing phishing datasets have one or more major problems that hamper replication by researchers.

\begin{enumerate}
    \item \emph{Dead.} Many phishing datasets contain lists of URLs. However many of these URLs are no longer functional which means that it is impossible to derive features and/or analyze the webpages associated with those URLs today. Examples of such datasets are PhishStorm and Ebbu.
    \item \emph{Feature Only.} Some datasets have the opposite problem: they include feature vectors associated with some websites but do not explicitly list those websites themselves. This means that defining new features and extracting them from the original webpage is impossible today. This is the problem with the Mendeley and UCI datasets.
    \item \emph{Non-Uniform Feature Sets.} Different datasets offer different sets of features. Having a uniform set of features across multiple datasets is necessary for fair evaluation of different PDs across different datasets---yet this turns out to be very difficult.
    \item\emph{Non-Replicability.} Some past efforts (e.g.,~\cite{tan2016phishwho,jain2018towards,niakanlahiji2018phishmon}) use public lists such as PhishTank or Alexa Top1Million\footnote{\url{https://www.alexa.com/topsites}} to build custom datasets. However, these lists are updated very often, hence it is not possible to replicate the exact same data used in those studies for comparison purposes. 
\end{enumerate}

Simply put, the four problems listed above make it impossible to test new approaches across these diverse datasets and show if/when the new approaches outperform the old ones. The major existing datasets either disclose the original URLs which are not active anymore, or they do not disclose the URLs in which case follow on research does not know where to look. In neither case can different features (even from existing papers) be added, let alone new features invented by follow on researchers.

\subsection{Solution: \LNU\ }
We address these issues by creating a new dataset, called \LNU\ (short for Liechtenstein and Northwestern University). With over $23\,000$ samples, \LNU\ is one of the biggest labeled datasets for phishing detection (the only bigger labeled datasets we know of are PhishStorm and Ebbu which suffer from the issues mentioned above). A comparison of \LNU\ with other datasets is given in Table~\ref{tab:datasets_comparison}.

\LNU\ is a \textit{large} and \textit{fixed} dataset that can be used for future work on PDs. It contains complete information on each sample, such as the \textit{URL}, the \textit{DNS} records, as well as the underlying \textit{HTML} code and a \textit{screenshot} of the webpage. In addition, the dataset includes the features listed in Table~\ref{tab:LNUfeatures}.  Hence, even if the webpage is taken down, the data in \LNU\ captures all information for reproducible future researchers; such information can also be augmented by, e.g., creating novel feature sets.

\begin{table}[ht!]
\centering
\caption{Comparison of \LNUscript\ with existing static datasets.}
\resizebox{\columnwidth}{!}{%
  \begin{tabular}{c||c|c|c|c|c|c|c|c}
  \toprule

  {\bf Name} &
  \textbf{Date} &
  {\begin{tabular}{@{}c@{}} \textbf{Phishing} \\ \textbf{samples}\end{tabular}} &
  {\begin{tabular}{@{}c@{}} \textbf{Legit} \\ \textbf{samples}\end{tabular}} &
  {\begin{tabular}{@{}c@{}} \textbf{URL} \\ \textbf{data}\end{tabular}} &
  {\begin{tabular}{@{}c@{}} \textbf{HTML} \\ \textbf{data}\end{tabular}} &
  {\begin{tabular}{@{}c@{}} \textbf{Reputation} \\ \textbf{data}\end{tabular}} &
  {\begin{tabular}{@{}c@{}} \textbf{Screenshot} \\ \textbf{data}\end{tabular}} &
  \textbf{Features} \\

  \midrule
  PhishLoad & 2012 & $3\,510$ & $8\,190$ & \cmark & \cmark & \xmark & \xmark & \xmark \\
  UCI & 2015 & $6\,050$ & $3\,950$ & \xmark & \xmark & \xmark & \xmark & \cmark \\
  DeltaPhish & 2017 & $1\,200$ & $4\,800$ & \cmark & \cmark & \xmark & \cmark & \xmark \\
  Ebbu & 2017 & $37\,175$ & $36\,400$ & \cmark & \xmark & \xmark & \xmark & \xmark \\
  PhishStorm & 2014 & $48\,000$ & $48\,000$ & \cmark & \xmark & \xmark & \xmark & \xmark \\
  Mendeley & 2018 & $5\,000$ & $5\,000$ & \xmark & \xmark & \xmark & \xmark & \cmark \\
  \midrule
  \LNUscript & 2020 & $7\,861$ & $15\,773$ & \cmark & \cmark & \cmark & \cmark & \cmark \\

  \bottomrule
  \end{tabular}
}
\label{tab:datasets_comparison}
\end{table}

\subsection{Creation Workflow of \LNU}
\label{sec:collection_lnu}
We collected benign samples from the Alexa Top-1million list and malicious samples from the well-known PhishTank and OpenPhish\footnote{\url{www.openphish.com}} repositories. All the entries were retrieved in March 2019.

To create a balanced corpus of benign websites, we divided the Alexa Top-1million list into three parts: the ``top'' partition includes websites from rank $1$ to $10\,000$; the ``middle'' partition includes websites ranked from $10\,001$ to $100\,000$; the ``bottom'' partition includes all websites ranked below $100\,001$. We extract $\sim5\,000$ websites from each partition. Our scripts visited each URL and saved the corresponding HTML as well as the full image representation of the homepage. We also queried and stored information provided by public DNSs for each URL.
To populate the phishing entries, we followed a procedure similar to that in~\cite{tian2018needle}. We monitored the PhishTank and OpenPhish sources for 3 weeks. Whenever a new phishing URL was added to these lists, we visited it and---if available----saved the HTML, the screenshot of the landing webpage, and the information provided by the DNS query.

The resulting $15\,773$ benign and $7\,861$ phishing samples represent the proposed \LNU\ dataset, which we publicly release at: \url{https://lnu-phish.github.io}.

\subsection{\LNU\ Dataset Features}
\label{sec:feature_lnu}

We compute the features (summarized in Table~\ref{tab:LNUfeatures}) for each sample in our \LNU\ dataset. The features are computed through the methodology in~\cite{basnet2014learning} and~\cite{mohammad2014predicting}. We focus on these features because they share many similarities with existing datasets and because they are used by several related efforts~\cite{tan2016phishwho,jain2018towards,niakanlahiji2018phishmon, mohammad2014predicting, abdelhamid2014phishing, verma2015character, jeeva2016intelligent}.
Using similar feature sets allows a more fair comparison of PDs devised over different datasets.\footnote{We are aware that other studies adopt more features (e.g.,~\cite{basnet2014learning, ma2009beyond}), but the non-reproducibility of the datasets used to validate their PDs does not allow us to compare our paper with their work.} 

\begin{table}[htpb!]
\centering
\caption{List of features included the \LNUscript\ dataset.}
\resizebox{0.8\columnwidth}{!}{%
        \begin{tabular}{|c|c|c|}
            \hline
            \textit{URL-features} & \textit{REP-features} & \textit{HTML-features} \\
            \hline
            \hline
            IP address & SSL final state & SFH  \\ \hline
            '@' (at) symbol & URL/DNS mismatch & Anchors \\ \hline
            '-' (dash) symbol & DNS Record & Favicon \\ \hline
            Dots number & Domain Age & iFrame  \\ \hline
            Fake HTTPS & PageRank & MailForm  \\ \hline
            URL Length & PortStatus & Pop-Up  \\ \hline
            Redirect & Redirections & RightClick \\ \hline
            Shortener &  & Objects \\ \hline
            dataURI &  & StatusBar \\ \hline
            &  & Meta-Scripts \\ \hline
            &  & CSS \\ \hline
        \end{tabular}
}
\label{tab:LNUfeatures}
\end{table}

Nevertheless, the information provided in our \LNU\ dataset allows the creation of any feature set for future works. In particular, we observe that ML-based detection systems must be periodically updated with recent data to avoid concept-drift problems~\cite{jordaney2017transcend, tian2018needle}. To further facilitate such `updates', we also release the source-code we developed to compute the features of \LNU, so that future researchers can `expand' it with more recent data by maintaining a uniform feature set.

%% file: sections/4-attacks.tex
\section{Proposed Gray Box Attacks on Phishing Detectors}
\label{sec:attacks}

We now describe the Gray Box attacks on Phishing Detectors considered in this paper, which can be divided into \textit{simple} and \textit{complex} attacks. In simple attacks, the attacker knows and targets only few specific features. However, in the complex attacks, we assume the attacker may know (and modify) a potentially huge number of the features used by the PD. We vary the percentage of features used by the defender that the attacker knows from 0 to 100\%. This range captures all adversarial scenarios: black box (no features known), white box (all features known), and gray box (some features known).

\subsection{Simple Attacks}
\label{sec:simple}
\begin{figure*}[!htbp]
    \centering
    \begin{subfigure}{0.97\columnwidth}
        \centering
        \frame{\includegraphics[width=\linewidth]{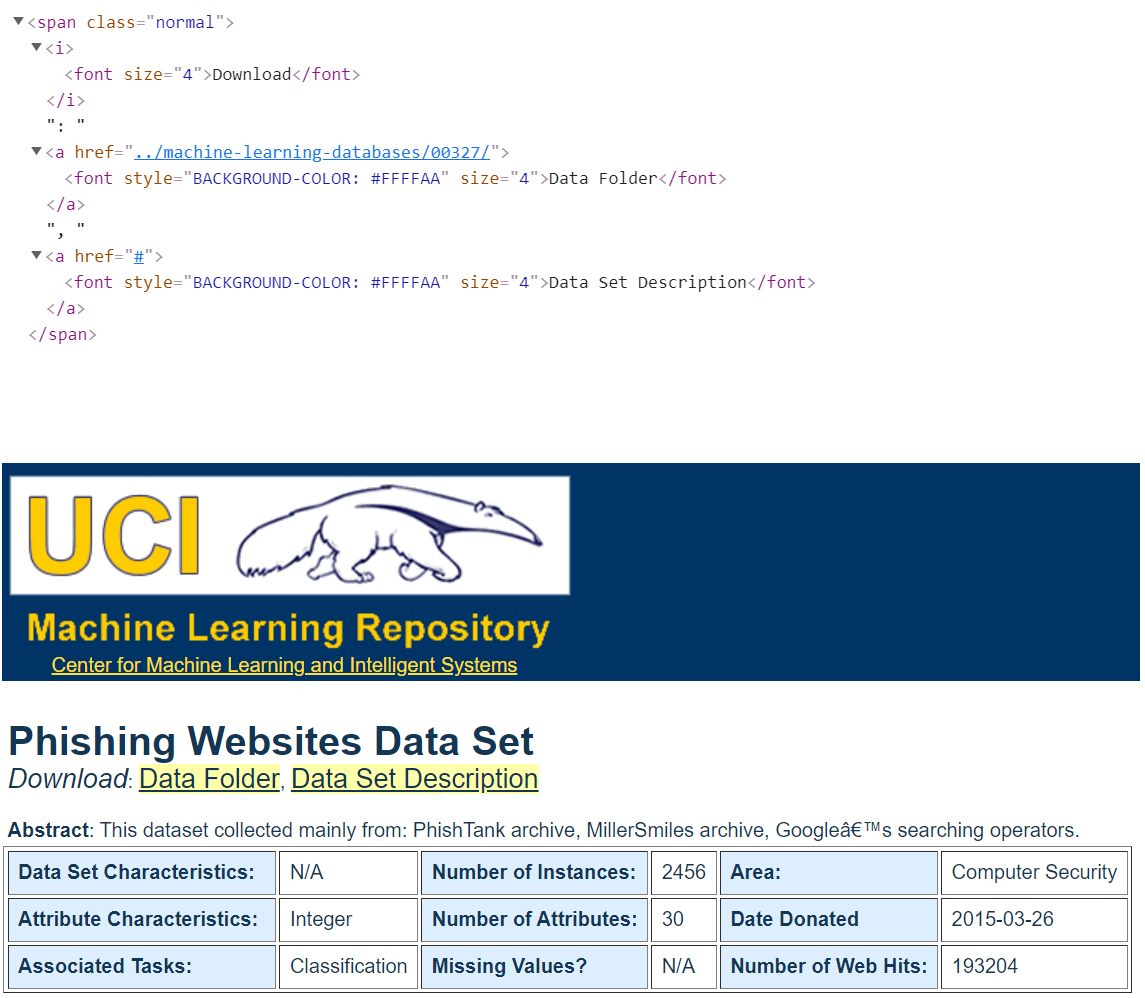}}
        \caption{The original webpage.}
        \label{sfig:gba2_original}
    \end{subfigure}\hfill%
    \begin{subfigure}{0.97\columnwidth}
        \centering
        \frame{\includegraphics[width=\linewidth]{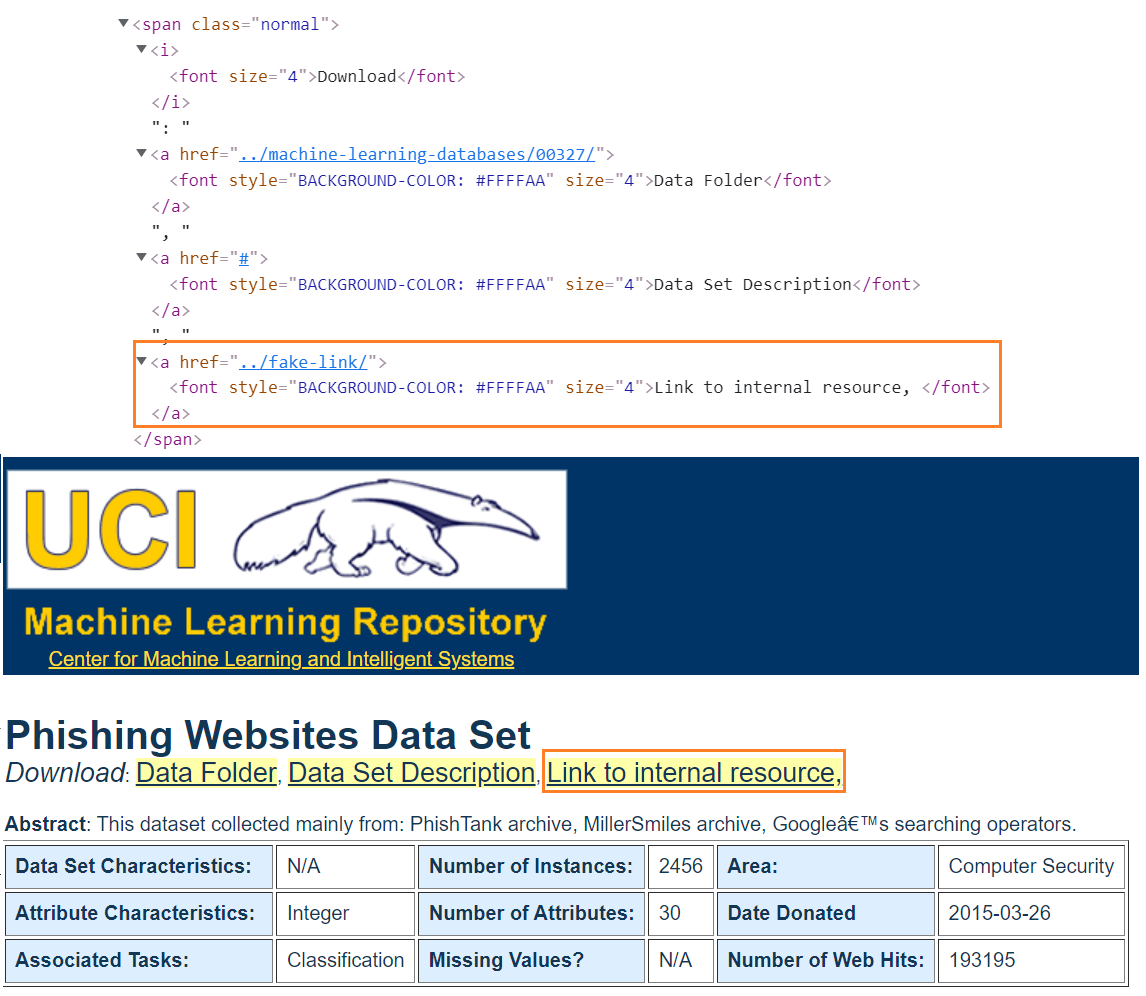}}
        \caption{The `adversarial' webpage.}
        \label{sfig:gba2_attack}
    \end{subfigure}\hfill%
    \caption{An example of the \GBA{2} attack. On the top there is the underlying HTML-code of the webpage, whereas at the bottom there is the rendered HTML shown in the webpage. The red box in Figure~\ref{sfig:gba2_attack} shows the manipulation to the HTML-code and its result to the real page. The `fake' link can also be hidden, preventing to be shown in the rendered HTML.}
    \label{fig:gba2}
\end{figure*}

We consider three very simple Gray Box attacks:

\begin{enumerate}
    \item \GBA{1}: The attacker assumes that the PD uses information about the length of the URL to predict whether it belongs to a phishing website or not---as is done in several existing PDs~\cite{babagoli2019heuristic, verma2015character, jeeva2016intelligent}. Hence it is reasonable to assume that an attacker may try to circumvent such mechanisms. Often times, phishing URLs are longer than benign URLs in order to confuse and trick users into clicking on the link. For this reason, existing PDs are usually trained on malicious samples characterized by longer URLs. Thus, an attacker may try to evade detection by devising phishing URLs with shorter URLs: a possible way to accomplish this is by using a URL shortening service (e.g., \url{tinyurl.com}).  
    \item \GBA{2}: Here, the attacker assumes that the PD uses features related to the HTML-code (this is done, for instance, in the PD considered in~\cite{subasi2017intelligent}), but he may not know exactly which feature. Hence, he tries to alter some aspects of his HTML code. For example, he might know that some PDs consider the ratio of internal links (i.e. within the URL's domain) to external links. Hence, in this attack, he inserts a number of ``internal'' links to his URL domain that might fool classifiers that use this feature. We inserted such internal links to a host of resources such as images, favicons, CSS snippets, videos, audio, as well as the usual ``textual'' links.
    Figures~\ref{fig:gba2} show how such an attack might be accomplished. The original webpage is in Figure~\ref{sfig:gba2_original}, whereas the adversarial webpage is in Figure~\ref{sfig:gba2_attack}. In particular, the top of Figures~\ref{fig:gba2} show the HTML-code, whereas the bottom show the rendered webpage. We can see from the red-box in Figure~\ref{sfig:gba2_attack} that by manipulating the HTML code it is possible to insert `fake' links to internal resources, which may favor an attacker to camouflage a phishing webpage into a benign webpage.
    \item \GBA{3}: This attack is a combination of the \GBA{1} and \GBA{2} attacks.
\end{enumerate}

Even though these attacks are relatively simple, there is considerable evidence that defenders use the HTML content in phishing websites and the structure of the URLs of phishing websites to build PDs~\cite{corona2017deltaphish,jain2018towards}. It is therefore reasonable to assume that attackers will try to use offensive techniques conforming to \GBA{1}--\GBA{3}.

\subsection{Complex attacks}
\label{sec:complex}
In our more sophisticated attacks \GBA{$\Delta$}, the adversary knows a variable subset of the features used by the defender. Let $\mathbb{F}_d$ be the set of features used by the defender (the PD) and $\mathbb{F}_a$ be the set of features that the attacker thinks the defender is using. This family of attacks is based on $\Delta$, the percentage of features actually used by the defender that the attacker guessed correctly, i.e. $\Delta = \frac{|\mathbb{F}_d\,\cap\, \mathbb{F}_a|}{|\mathbb{F}_d|}.$
In \GBA{$\Delta$} attacks, we vary $\Delta$ by assuming the attacker knows some $\mathbb{F}_a$.

The 27 basic features used (cf. Table~\ref{tab:LNUfeatures})  can be easily manipulated by experts who can easily insert/remove redirections, synthetically modify the URL length, or change any HTML functionality of the webpage to alter the features shown in Table~\ref{tab:LNUfeatures}. This enables launching a huge number of attacks using \GBA{$\Delta$}.

Note that when $\Delta=0$, we have a black box attack, and when $\Delta=1$, we have a white box attack. 

In this paper, we evaluate 7 different values of $\Delta$. Hence \GBA{$\Delta$} represents many attack scenarios (7 in our case) where the adversary is able to modify any selection of $\Delta$\% of the total set of features considered. This leads to an huge number of possibilities, viz. $2^{\Delta * |\mathbb{F}_d|}\text{--}1$ which is a very large space of possible attacks.
Simply put, \GBA{$\Delta$} captures many possible complex evasion attacks conceivable by well-motivated and expert opponents.


\subsection{Impact of Adversarial Attacks}
\label{sec:impact}
In this paper, we consider attacks on 13 well known classifiers.
9 are classical classifiers:
Random Forest (RF),
K-Nearest Neighbor (KNN),
Decision Tree (DT),
Logistic Regression (LR),
Naive Bayes (NB), 
Support Vector Machines (SVM), 
Extra Trees (ET), 
Stochastic Gradient Descent (SGD), 
Bagging (Bag).
We also consider 2 boosting techniques---AdaBoost (AB) and Gradient Boost (GB). Finally, we also evaluate 2 deep learning classifiers: Multi-Layer Perceptrons (MLP) and Google's recent ``Deep and Wide'' (DnW) method~\cite{cheng2016wide}.
\emph{Thus, we note that our Gray Box attackers assume that the PD is using any one of these 13 classifiers, but they do not need to know which one.}
We report an overview of existing PDs and the datasets used for their evaluation in Table~\ref{tab:past-pds}, from which we observe that many of our classifiers are used by related work.

\begin{table}[htpb!]
\centering
\caption{Classifiers and Dataset of existing PDs.}
\resizebox{0.7\columnwidth}{!}{%
        \begin{tabular}{|c|c|c|}
            \hline
            \textit{Reference} & \textit{Classifier} & \textit{Dataset} \\ \hline
            \hline
            ~\cite{mohammad2014predicting} & MLP & Custom \\ \hline 
            ~\cite{basnet2014learning} & RF, NB, MLP, LR, SVM & Custom \\ \hline 
            ~\cite{basnet2015towards} & RF, SVM, NB, MLP, LR & Custom \\ \hline 
            ~\cite{verma2015character} & RF, NB, LR & Custom \\ \hline 
            ~\cite{abdelhamid2017phishing} & RF, SVM, AB & UCI \\ \hline 
            ~\cite{subasi2017intelligent} & RF, KNN, SVM, MLP, NB & UCI \\ \hline 
            ~\cite{ali2017phishing} & MLP, SVM, KNN, NB, RF & UCI \\ \hline 
            ~\cite{corona2017deltaphish} & SVM & DeltaPhish \\ \hline 
            ~\cite{babagoli2019heuristic} & LR, SVM & UCI \\ \hline 
            ~\cite{jain2018towards} & RF & Custom \\ \hline 
            ~\cite{niakanlahiji2018phishmon} & RF, KNN, AB & Custom \\ \hline 
            ~\cite{sahingoz2019machine} & RF & Ebbu \\ \hline 

        \end{tabular}
}
\label{tab:past-pds}
\end{table}

We define the \emph{Impact} of attack $Att^d_{i}$ of type $i$ on dataset $d$ and classifier $Clf$ on a performance metric $\mu$ with the following Equation:
\begin{equation}
\resizebox{0.9\columnwidth}{!}{
$Impact(Att^d_i,Clf,\mu) =
\frac{\mu(Clf|\neg Att^d_i)-\mu(Clf|Att^d_i)}{\mu(Clf|\neg Att^d_i)}$
}
\label{eq:impact}
\end{equation}
where $\mu(Clf|Cond)$ denotes the value of the performance metric $\mu$ of classifier $Clf$ when condition $Cond$ is true.\footnote{We consider two conditions: if an attack is present or absent.} In the above formulation, $\mu$ can be any measure of classifier performance (e.g. F1-score, Accuracy, etc).

For instance, suppose the F1-score of a given classifier (say Random Forest) is 80\% when no attack is performed, and 60\% when an attack is launched on it. In this case,
$Impact(Att^d_i,RF,F1)=\frac{0.8-0.6}{0.8}=0.25$, i.e. there is a 25\% reduction in the F1-score. Note that Equation~\ref{eq:impact} can be used to measure the impact of \textit{any} adversarial attack, not just Gray Box attacks.

We expect that \textit{all} the considered Gray Box attacks (\GBA{1}--\GBA{$\Delta$}) have a significant impact on classification algorithms used in the literature~\cite{abdelhamid2014phishing,verma2015character,jeeva2016intelligent,abdelhamid2017phishing,corona2017deltaphish,babagoli2019heuristic,jain2018towards,sahingoz2019machine}.
But before we evaluate this, we introduce our proposed defensive mechanism.

%% file: sections/5-defense.tex
\section{Proposed Countermeasure: The \POC\ Algorithm}
\label{sec:defense}
In this section, we introduce the concept of \emph{operation chains} or $oc$, and the \POC\ algorithm. The basic idea behind \POC\ is to create a new feature space ($\Psi$) by randomly mixing some of the features from $\mathbb{F}$ used by a given PD. This makes it hard for the attacker to infer \textit{which} features are used by a PD and \textit{how} such features denote a phishing/benign webpage. In particular, even if the attacker knows $\Delta$\% of the features of the PD, he may not know how to change them in order to make a malicious webpage be classified as benign. To achieve this, we use operation chains which consist of a base set of mapping operators that randomly transform some features into a new feature, which will be included in the actual feature space used by the `hardened' PD.
Simply put, the \POC\ algorithm  obfuscates the features used by a PD so that an adversary cannot easily tell what the PD is doing\footnote{Our work is different from the type of obfuscation that a might perform in  order to stop his/her code from being reverse engineered. But the idea is similar in both cases.} and, hence, offensively react to such PD.

\subsection{Formal description of \POC}
\label{ssec:description}
Without loss of generality, \POC\ assumes that the features are numeric (real valued) and that categorical feature values will be replaced by values from a discrete domain.
The obfuscation provided by \POC\ is done via certain kinds of mappings. 

A \emph{unary (resp. binary) feature mapping} operator $\alpha$ (resp. $\beta$) is a mapping from $\mathbb{R}$ (resp. $\mathbb{R}\times \mathbb{R}$) to $\mathbb{R}$. We assume the existence of a set $\Fmop$ of unary and binary feature mapping operators. 
To obfuscate the underlying mechanism of \POC, the $\Fmop$ set should consist of mappings that are hard to reverse. There are many such operations of course, and so we choose a suite of well-known, nonlinear mappings.
In our implementation, we use $\Fmop\!\!=\!\! \{log,sin,cos,tan,exp^i,+,-,*,/\}$ as our feature mapping operators where $\{log,sin,cos,tan\}$ are unary operators, $exp^i$ is not one but a family of numeric unary operators which take an input value $x$ and return $x^i$ for $i\in\!\{\text{-}3,\text{-}2,\ldots,2,3\}$. The arithmetic operators $\{+,-,*,/\}$ are binary feature mapping operators\footnote{Again, please note that these are the $i$'s used in our implementation. The theory allows $i$ to range over any set of integers.}. Note that our definitions below apply to virtually any choice of unary and binary operators in $\Fmop$---we are not limited in any way to the specific operators chosen in our implementation---new ones and the definition of operation chains below can be seamlessly incorporated into our framework\footnote{We do not claim that our selected mappings are the best. There is an infinite space of such mappings which can be used in \POCscript.}.

The \POC\ algorithm maps a given set of features $\mathbb{F}$ into a new feature space $\Psi$, composed of operation chains, $oc$. Each $oc$ is based on a subset $\overbar{\mathbb{F}}\!\subseteq\!\mathbb{F}$ of features, which are combined via the unary or binary operators in $\Fmop$. The final $\Psi$ is then created by randomly selecting $\psi$ (representing the dimensionality of $\Psi$) $oc$. Specifically, given a set $\mathbb{F}$ of (original) features, we can recursively define \emph{operation chains}, each based on $\overbar{\mathbb{F}}\!\subseteq\!\mathbb{F}$, as follows:
\begin{enumerate}
    \item Each feature $f\!\in\!\overbar{\mathbb{F}}$ is an $oc$ of size $0$.
    \item For each unary operator $\alpha\!\in\!\Fmop$ and for each $oc$ of size $s$, $\alpha(oc)$ is an $oc$ of size $s\!+\!1$.
    \item For each binary operator $\beta\!\in\!\Fmop$ and for each pair of operator chains $oc_1,oc_2$ of sizes $s_1,s_2$ respectively, $\beta(oc_1,oc_2)$ is an $oc$ of size $s_1\!+\!s_2\!+\!1$.
\end{enumerate}
Suppose $\mathbb{F}$ is the list of features shown in Table~\ref{tab:LNUfeatures} and suppose $\overbar{\mathbb{F}}$ consists of any two of these features, e.g. $\overbar{\mathbb{F}}=\{f_1,f_2\}$. Then examples of operation chains based on $\overbar{\mathbb{F}}$ and on the proposed set of $\Fmop$ include:
\begin{enumerate}
    \item $f_1$ and $f_2$ are both $oc$ of size $0$.
    \item $\mbox{sin}(f_1), \mbox{cos}(f_2)$ are $oc$ that create new features by taking the sine and cosine, respectively of values of features $f_1,f_2$ respectively. They have size $1$.
    \item $\mbox{sin}(f_1)\!+\!\mbox{cos}(f_2)$ is an $oc$ that generates a new feature that creates feature values by summing up the sine of the value of feature $f_1$ and the cosine of the value of feature $f_2$. This $oc$ has size $3$.
    \item $exp(\mbox{sin}(f_1)\!+\!\mbox{cos}(f_2))$ creates a new feature whose value is $e^{\mbox{sin}(f_1)+\mbox{cos}(f_2)}$. The size of this $oc$ is $4$.
\end{enumerate}

\noindent
Thus our \POC\ framework creates $\Psi$ as follows.  
\begin{enumerate}
    \item (Initialization) 
        First, we select $\overbar{\mathbb{F}}$ from $\mathbb{F}$, representing the features used to create each $oc$. We then define the set of mapping operators, $\Fmop$, and choose $MaxSize$ which is an integer greater than 0; finally, we set $\psi$, representing the cardinality of the new feature space $\Psi$.
    \item (Operation Chain Transformations) 
        We create $oc$ of size $MaxSize$ or less by combining the features in $\overbar{\mathbb{F}}$ via the operators in $\Fmop$.
    \item (Random Selection) 
        We randomly select $\psi$ operation chains, which will represent $\Psi$.
\end{enumerate}

The \POC\ Algorithm that formally captures the informal process described above is shown in Algorithm~\ref{alg:poc}.

\input{sections/algorithms/generateFeatureMapping.tex}

\subsection{Analysis of \POC}
\label{ssec:analysis}
We now analyze our \POC\ algorithm.

\textbf{Relationship between $\Psi$ and $\mathbb{F}$.}
We note that each feature in $\Psi$ is represented by an $oc$ that uses a subset of the features in $\mathbb{F}$. Hence, the new $\Psi$ and the original $\mathbb{F}$ can be linked by how many features of $\mathbb{F}$ are included in $\Psi$. Let us define the \textit{prevalence} of $\mathbb{F}$ w.r.t. a given $\Psi$ as $\mathcal{P}(\mathbb{F},\Psi)$, which denotes
the percentage of features in $\mathbb{F}$ that are included among all $oc$ composing $\Psi$. As an example, if $\mathbb{F}\!=\!(f_1, f_2, f_3, f_4)$ and $\Psi\!=(oc_1, oc_2)$ with $oc_1\!=\!(f_1\!+\!f_2)$ and $oc_2\!=\!(sin(f_3)\!+\!f_1)$ then $\Psi$ contains three out of four features of $\mathbb{F}$ (specifically, $f_1, f_2, f_3$), meaning that $\mathcal{P}(\mathbb{F},\Psi)\!\!=\!\!75\%$.
Two cases are possible:
\begin{itemize}
    \item (complete \textit{prevalence)} $\mathcal{P}(\mathbb{F},\Psi)\!=\!100\%$, i.e., $\Psi$ uses \textit{all} the features in $\mathbb{F}$. In this case\footnote{Of course, this can only be true if $\overbar{\mathbb{F}}=\mathbb{F}$.}, we can expect that using \POC\ results in a PD with similar performance \textit{in the absence of adversarial attacks} as a PD that does not use \POC, because they will both use the same amount of information available to analyze each sample.
    \item (incomplete \textit{prevalence}) $\mathcal{P}(\mathbb{F},\Psi)\!<\!100\%$, i.e., $\Psi$ does not contain some features of $\mathbb{F}$. In this case, using \POC\ will result in PDs that are trained on less information (due to the `excluded' features), but with the capability of completely nullifying those adversarial attacks that target features of $\mathbb{F}$ not included in the $oc$ of $\Psi$ (by leveraging the well-known \textit{feature removal} strategy~\cite{smutz2012malicious, apruzzese2019addressing}).
\end{itemize}
We will investigate both of these circumstances.

\vspace{0.5em}

\textbf{Goal of \POC.}
\POC\ assumes that attackers can only change some features.\footnote{The assumption is realistic, because some features cannot be changed without altering the malicious nature of a webpage, or require a huge resource investment (e.g., modifying reputation features based on DNS records requires compromising the respective DNS servers).} 
Hence, \POC\ seeks to prevent  hackers from: (i)~reverse-engineering the PD by identifying its complete feature set; and (ii)~changing one or two small things to evade a PD.

The first goal is achieved by randomly using the old features ($\mathbb{F}$) to create a new feature space ($\Psi$) which makes it harder to reverse-engineer (or `steal'~\cite{wang2018stealing}) the classifier used by the PD (e.g. the attack against the Google Chrome filter~\cite{liang2016cracking}). This is because the attacker's manipulation will affect multiple features simultaneously and  differently, making it hard for him to infer the features used by the PD. Moreover, as described in Section~\ref{sec:feature_lnu}, ML detectors must be updated with new data to prevent concept drift~\cite{jordaney2017transcend, tian2018needle}. Therefore, each new application of \POC\ will result in a new feature space, ensuring that attackers that `cracked' the old PDs have to repeat the process again.

The second goal is achieved as a direct consequence of the above. The new feature space $\Psi$ induces confusion, e.g. a feature $f \in \mathbb{F}$ may be mapped to $sin(f)$ and then further combined into an $oc$ such as $2^{sin(f)+log(f)}$. By using irreversible feature mappings, it is unlikely that the attacker can recover the original features---even if the attacker were to get hold of the \POC-feature vectors describing each sample. Hence, an attacker may be successful in evading a PD by ``making a URL shorter'' (as in \GBA{1}), thus manipulating the corresponding feature. However, against a \POC-hardened PD, the manipulated feature will affect many $oc$s in unpredictable ways. Increased protection is also provided by feature-removal (cf. Section~\ref{ssec:analysis})---e.g., if the features manipulated in \GBA{1} are not used (or nullified\footnote{Feature removal can occur `indirectly' even if $\mathcal{P}(\mathbb{F},\Psi)\!\!=\!\!100\%$, e.g., if $\Psi$ contains an $oc$ where a feature is multiplied by another feature whose value is 0 for most samples.}) by a \POC-hardened PD. Nevertheless, as operation chains get longer, recovering the original features from the new feature values is very challenging. Therefore, the selected mapping operators (e.g. \textit{log}, \textit{sin}, \textit{cos} used in this paper) should include many irreversible functions that make the attacker's job even more difficult.

\textbf{Summary.} As long as $\Psi$ contains enough of the original features (e.g., $\mathcal{P}(\mathbb{F},\Psi)\!\!\geq\!\!70\%$), and as long as such $\Psi$ has high dimensionality (e.g., $\psi\!\geq\!15 oc$), a \POC-hardened PD has the potential to: capture enough of the distributional properties of the original data (providing good classification power); obfuscating the features of the PD (confusing the attacker); and mitigating the attacks. Finally, we stress that the mappings we use in this paper can be replaced (or expanded) by other mappings (e.g. hyperbolic tangent) that are irreversible. 
Our website also contains our implementation of \POC.

%% file: sections/algorithms/generateFeatureMapping.tex
\SetKwRepeat{Do}{do}{while}%
\SetKw{KwBy}{by}%
\begin{algorithm2e}
    \caption{Proposed \POC\ algorithm.}
    \footnotesize
    \label{alg:poc}
    \DontPrintSemicolon
    \SetAlgoNoEnd
    
    \KwIn{List of features $\overbar{\mathbb{F}}$ included in a given dataset $d$; $\phi$, number of new features that will compose the new set of $mapped\_features$; $MaxSize$ maximum size of an operation chain; $\Fmop_\alpha$ set of unary feature mapping operators; $\Fmop_\beta$ set of binary feature mapping operators}
    \KwOut{The {\scriptsize$mapped\_features$} representing the new space $\Phi$.}
    
    \hrule

    $mapped\_features \gets$ emptyList();\\
    \For{$h \gets0$ \KwTo $\phi$ \KwBy $1$}{
        $new\_feature \gets$ computeNewFeature($\overbar{\mathbb{F}}$, $MaxSize$);\\
        Insert $new\_feature$ in $mapped\_features$;\\
        }
    \Return $mapped\_features$
    
    {\tt \scriptsize // Procedure that generates a $mapped\_feature$}\\

    \SetKwFunction{proc}{proc}
    \SetKwProg{myproc}{Function}{}{}
    \myproc{computeNewFeature($\overbar{\mathbb{F}}$, $MaxSize$)}{ 
    \label{alg:computeNewFeature}
        $feature\_block \gets$ chooseFeatures($\overbar{\mathbb{F}}$, $MaxSize$);\\
        $ L \gets$ len($feature\_block$); \\
        \For{$h\gets0$ \KwTo L \KwBy $1$}{
            $\overbar{f}\gets$ unaryOperation($feature\_block[h]$);\\
            Replace $feature\_block[h]$ with $\overbar{f}$;\\
        }
        \For{$h\gets1$ \KwTo L \KwBy $1$}{
            $\overbar{f}\gets$ binaryOperation($feature\_block[0]$,$feature\_block[h]$); \\
            Replace $feature\_block[0]$ with $\overbar{f}$; \\
        }
        $new\_feature\gets feature\_block[0]$;\\
        \Return $new\_feature$
    }

    {\tt \scriptsize // Procedure that determines the features from $\overbar{\mathbb{F}}$ considered to generate a $mapped\_feature$}\\

    \SetKwFunction{proc}{proc}
    \SetKwProg{myproc}{Function}{}{}
    \myproc{chooseFeatures($\overbar{\mathbb{F}}$, $MaxSize$)}{ 
    \label{alg:chooseFeatures}
        $feature\_block\gets$emptyList();\\
        \For{$h\gets0$ \KwTo randomChoice($MaxSize$) \KwBy $1$}{
            Insert randomChoice($\overbar{\mathbb{F}}$) in $feature\_block$;\\
        }
        \Return $feature\_block$
    }
    
    {\tt \scriptsize // Procedure that chooses and applies an unary operation among $Fmop_\alpha$ on a given feature $f$ that composes a given $new\_feature$.}\\
    
    \SetKwFunction{proc}{proc}
    \SetKwProg{myproc}{Function}{}{}
    \myproc{unaryOperation($f$)}{ \label{alg:unaryOperation}
        $\overbar{f}$ $\gets$ Apply $randomChoice(\Fmop_\alpha)$ to $f$; \\
        \Return $\overbar{f}$\\
    }
    
    {\tt \scriptsize // Procedure that chooses and applies a binary operation among $Fmop_\beta$ on a given pair of features ($f_1$, $f_2$) that composes a given $new\_feature$.}\\
    
    \SetKwFunction{proc}{proc}
    \SetKwProg{myproc}{Function}{}{}
    \myproc{binaryOperation($f_1$, $f_2$)}{ \label{alg:binaryOperation}
        $\overbar{f}$ $\gets$ Apply $randomChoice(\Fmop_\beta)$ to $f_1$ and $f_2$; \\
        \Return $\overbar{f}$\\
    }

\end{algorithm2e}

%% file: sections/6-experiments.tex
\section{Experiments}
\label{sec:experiments}
We use Python3 code (and leverage Scikit-Learn) to extract all the features in Table~\ref{tab:LNUfeatures}, as well as all the Gray Box attacks and the \POC\ algorithm. 

Our experiments focus on validating \POC\ w.r.t. (i)~mitigating Gray Box attacks and (ii)~maintaining high performance when there are no attacks. For this, we first assess the performance of existing PDs and their \POC\ versions in the absence of attacks. This shows how \POC\ performs when no attacks occur, and establishes the performance of the selected PDs.
We then evaluate the \textit{Impact} of the attacks on existing PDs and their \POC\ versions. We first show performance w.r.t. \textit{simple} attacks (\GBA{1} to \GBA{3}), and then show the results against \textit{complex} attacks (the 7 variants of \GBA{$\Delta$}). All these evaluations involve the 13 classifiers trained on each of the 4 datasets considered in the paper.
We discuss all these results in Section~\ref{sec:discussion}.

The rest of this section assesses \POC\ with complete \textit{prevalence}, i.e., when $\mathcal{P}(\mathbb{F},\!\Psi)\!\!=\!\!100\%$. The motivation is twofold: ensure a fair comparison of the performance between the baseline and \POC-hardened classifiers; and assess the hardening of \POC\ provided by its (random) feature mapping, and not due to the exclusion of the `attacked' features (cf. Section~\ref{ssec:analysis}). We will also evaluate \POC\ when $\mathcal{P}(\mathbb{F},\Psi)\!<\!100\%$ in Section~\ref{sec:prevalence}.

\subsection{Testbed}
\label{ssec:testbed}

\textbf{Baselines.} Our experimental testbed contains 13 `baseline' classifiers (used in existing phishing detectors) and the 4 different datasets considered in this paper (UCI, Mendeley, DeltaPhish, \LNU---see Section~\ref{sec:LNU}). 
For each dataset, all such baseline classifiers adopt the same set of features. For the \LNU\ and DeltaPhish datasets, we use the features in Table~\ref{tab:LNUfeatures} (we manually compute the \textit{REP-}features in the DeltaPhish dataset using the same methodology adopted to create our \LNU\ dataset---see Section~\ref{sec:LNU}). For the other datasets (Mendeley, UCI), we use all the features provided by their creators. We note that all these datasets share similar \textit{URL-} and \textit{HTML-}features, which are all affected by the attacks considered in this paper. We apply an 80:20 split for the training and test partitions, where the proportion of benign-to-malicious samples is as reported in Table~\ref{tab:datasets_comparison}. We did extensive hyper-parameter optimization using grid search across the parameter space of each classifier in order to tune that classifier to achieve best performance. The results show the performance after this hyper-parameter optimization (the most influential parameters for our algorithms are provided in the supplementary material).

\textbf{\POC-hardened classifiers.} We consider the (fine-tuned) `baseline' variant of each classifier as basis. We use the $\Fmop$ in Section~\ref{ssec:description}, and specify $\overbar{\mathbb{F}}\!\!=\!\!\mathbb{F}$, $MaxSize\!\!=\!\!3$, and $\psi\!\!=\!\!20$, meaning that $\Psi$ is composed of 20 $oc$. Because we are considering $\mathcal{P}(\mathbb{F},\Psi)\!\!=\!\!100\%$, the \POC\ classifiers include---across their 20 $oc$---all features of the baseline classifiers (yielding a `pseudo-random' \POC). 
We train (and test) each \POC\ classifier on all the datasets considered in the paper by using the same splits. Each classifier (on each dataset) adopts the mapping produced by \POC\ that achieves the best performance during development (i.e., in the no-attack case): in reality only a single PD (i.e., the one with best performance) is deployed, and the adversarial attacks cannot be anticipated. 
The Gray Box attacks are generated by using the malicious samples in the test partition as base. When we compute the \textit{Impact} of an attack, we rely on the Recall as performance metric (see Equation~\ref{eq:impact}), because we consider \textit{evasion} attacks which involve only malicious samples. The 7 attacks of \GBA{$\Delta$} family are repeated 10 times, each time by modifying different features (but always corresponding to the same $\Delta$), and the reported values correspond to the average of these 10 trials.

\POC\ is not likely to
yield good results when used with randomly chosen parameters. During training time, we identify the best parameter settings for each classifier using standard grid-search based hyper-parameter optimization. \footnote{Grid search looks at all the possible combinations of $\mathbb{F}$ that result in $\Psi$. To make this humanly feasible, we considered 100 combinations and choose the best one for each classifier.} Once the classifier is trained, the resulting \POC\ `hardened' classifier obtains a performance comparable to the `baseline' (in the absence of adversarial attacks) on test data not seen during training. Though costly, training is a one time operation. We will discuss the run-time for training in Section~\ref{ssec:time}.

\subsection{No Attack Case: Performance of Baselines vs. \POC}
\label{sec:no_attack}
We first assess all PDs (the `baselines' and their \POC-hardened variants) when no adversarial attacks occur.

\begin{figure}[!htbp]
    \centering
    \begin{subfigure}{0.5\columnwidth}
        \centering
        \includegraphics[width=\linewidth]{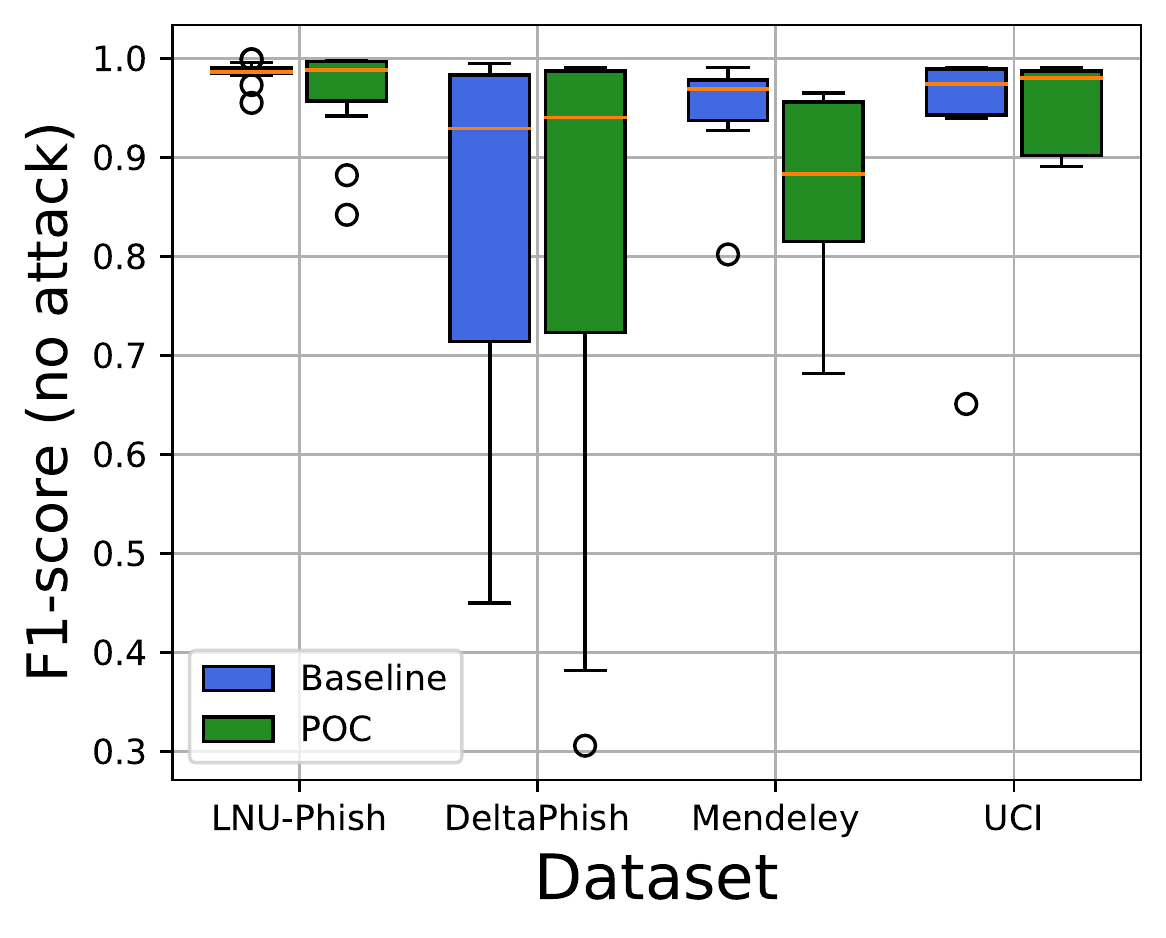}
        \caption{Distribution of the F1-score.}
        \label{sfig:no-attack_f1}
    \end{subfigure}\hfill%
    \begin{subfigure}{0.5\columnwidth}
        \centering
        \includegraphics[width=\linewidth]{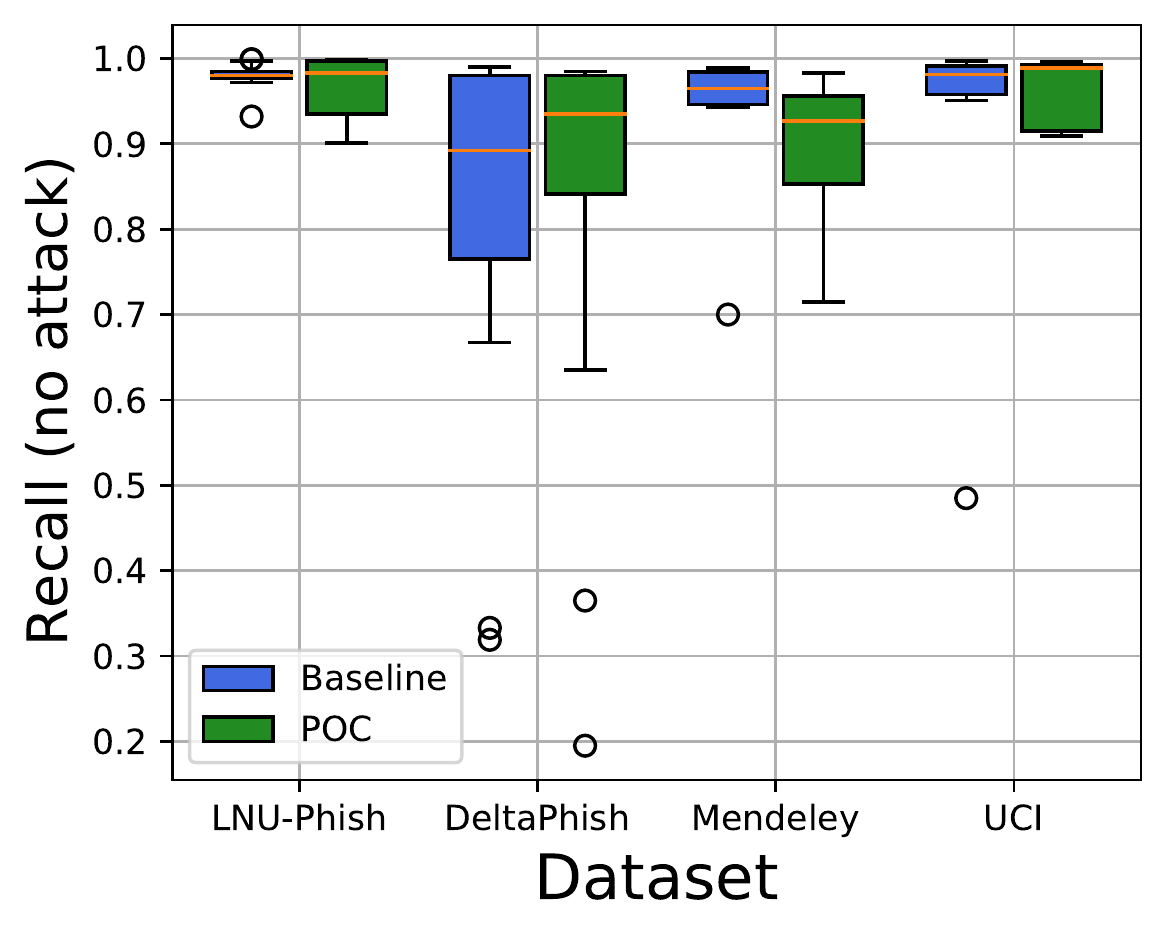}
        \caption{Distribution of the Recall.}        \label{sfig:no-attack_recall}
    \end{subfigure}\hfill%
    \caption{Comparison (baseline vs \POCscript) in the no-attack case.}
    \label{fig:boxplot_no-attack}
\end{figure}

Table~\ref{tab:expt-baseline2} shows the result of using the 13 baseline classifiers on the four datasets as done by past work, while Table~\ref{tab:expt-trans2} shows the result of using the \POC\ approach using all the features in the dataset. We tested the efficacy of \POC\ in the no attack case because of concerns that the random manipulation of features could lead to a drop in performance.
\nop{Table~\ref{tab:expt-baseline2} shows that our implementation of some baseline classifiers achieve similar performance as past work: for example,~\cite{sharma2020feature} achieves 96\% F1-score on the UCI and Mendeley Phishing datasets, whereas~\cite{deltaphish2018dataset} obtains 95\% Accuracy on the DeltaPhish dataset; these results are matched by our RF classifiers evaluated on these datasets.}
By comparing Table~\ref{tab:expt-baseline2} with Table~\ref{tab:expt-trans2}, we observe that the notion of operation chains used by \POC\ does lead to a small reduction in performance of the best classifier in each case (see the last row of Tables~\ref{tab:expt-baseline2} and~\ref{tab:expt-trans2}), but we note that this drop is very small (we will discuss such results in Section~\ref{ssec:statistical}). 
To better visualize the magnitude of the drop,  Figures~\ref{fig:boxplot_no-attack} shows the distribution of the F1-score (Figure~\ref{sfig:no-attack_f1}) and Recall (Figure~\ref{sfig:no-attack_recall}) achieved by the baseline classifiers (blue boxplots) and their \POC\ variant (green boxplots) on each dataset. We see that some nontrivial degradation only occurs with the Mendeley dataset, but a close look at Table~\ref{tab:expt-trans2} reveals that the best classifiers still yield high performance in the absence of attacks: for instance, the baseline ET classifier on Mendeley has 0.99 F1-score and Recall, and its hardened \POC\ variant has 0.96 F1-score and 0.95 Recall.

However, the next few experiments show that \POC\ outperforms past work when the adversary carries out any of the considered attacks (both simple and complex), and hence this negligible reduction in performance is amply compensated by \POC's increased robustness.

\input{sections/tables_big/noAtk_tables}

\subsection{Attacks against existing PDs}
\label{sec:evaluation_existing}

We assess the \textit{Impact} of the considered attacks on the baseline classifiers. We begin with the simple attacks (\GBA{1}, \GBA{2} and \GBA{3}) and then proceed with the complex attacks (the 7 variants of \GBA{$\Delta$}). 

\paragraph*{Impact of simple attacks}
Table~\ref{tab:gba-impact} shows the \textit{Impact} of the \GBA{1}--\GBA{3} attack on each of the 4 datasets and 13 classifiers (used in existing phishing detectors). We see, for instance, that \GBA{1} against the RF of the \LNU\ dataset leads to a $12.4\%$ drop, but the drop is $96.6\%$ on the DeltaPhish dataset. On average (last rows of subtables in Table~\ref{tab:gba-impact}), the \GBA{1}--\GBA{3} attacks lead to significant drops on all datasets (varying from $11.7\%$ to $90.2\%$), irrespective of the classifier used.

\paragraph*{Impact of complex attacks}
We now consider the complex attacks represented by \GBA{$\Delta$}, which assume that the attacker knows $\Delta$\% of the features used by the targeted PD. We vary $\Delta$ from $10-70\%$ in steps of $10\%$. 
Table~\ref{tab:gba4-impact} shows the \textit{Impact} of these attacks on existing PDs on all 4 datasets.  Unsurprisingly, as $\Delta$ increases, the attacks have a greater \textit{Impact} on average (the last line showing ``averages'' in the 4 subtables of Table~\ref{tab:gba4-impact} showing steady increases from left to right). Moreover, some of the attacks are very effective---for instance, if the attacker knows $30\%$ of the features used by the defender, the \textit{Impact} ranges from $15.7\%$ to $43.1\%$ which is very substantial. A formal statistical analysis of such results in presented in Section~\ref{ssec:statistical}.

\input{sections/tables_big/impact-Base_tables}

We observe that some attacks caused a negative \textit{Impact} (e.g., the NB in Table~\ref{tab:gba4-impact_Mendeley}), implying that the PD was able to correctly recognize \textit{more} phishing samples than in the absence of attacks. Such occurrence is a byproduct of a less than optimal training phase, because the adversarial manipulation resulted in a sample that the classifier considers to be ``more malicious'' than its non-modified variant (as shown in Table~\ref{tab:expt-baseline2}, the NB classifier on the Mendeley dataset achieves the lowest performance).

\subsection{Attacks against \POC}
\label{sec:evaluation_POC}
We now assess the effectiveness of \POC\ in protecting against the considered Gray Box attacks. We do so by measuring the \textit{Impact} of every attack against the \POC\ version of each baseline classifier, and computing the \textit{Impact difference} between the baseline and its \POC\ variant. This allows an immediate understanding of the results: if the number is greater than $0$, then \POC\ mitigated the attack; otherwise, it was more affected.

We begin by evaluating the simple attacks, and then conclude with the complex attacks.

\paragraph*{Simple attacks}
We assess \POC\ against the simple \GBA{1}--\GBA{3} attacks. These results are reported in Table~\ref{tab:gba-difference}. A positive difference (shown in bold) means that \POC\ was more resilient to the attack than the baselines, while a negative number means \POC\ was less resilient; higher values are highlighted with a darker background. Table~\ref{tab:gba-difference} consists mostly of bold entries, showing that \POC\ is more resilient than past work for almost all combinations of dataset and classifier used.
Additionally, we see from the last rows (``average'') that on average \POC\ exhibited superior performance for each of the 4 datasets considered: the \textit{Impact} of the \GBA{1}--\GBA{3} attacks on the baselines are $1\%$ to $53.5\%$ higher than for \POC\ (last row of the subtables in Table~\ref{tab:gba-difference}).

\paragraph*{Complex attacks}
We now turn to the value of \POC\ in protecting against the 7 variants of the \GBA{$\Delta$} attacks. The results are reported in Table~\ref{tab:gba4-difference}. A positive difference (denoted in bold) means that \POC\ was more resilient to the attack than the baselines, while a negative number means \POC\ was less resilient; higher values are highlighted with a darker background. We see that most entries in the table are in boldface, suggesting that \POC\ is more resilient to the \GBA{$\Delta$} attack irrespective of the dataset and classifier used.
As can be seen from the last rows (``average''), \POC\ exhibited superior performance for 27 of 28 combinations of dataset and classifier; the one exception is the UCI dataset with $\Delta=60$\% where the performance of the baseline is very slightly better than that of \POC.

\input{sections/tables_big/impact-POC_tables}

\begin{figure}
    \centering
    \includegraphics[width=\columnwidth]{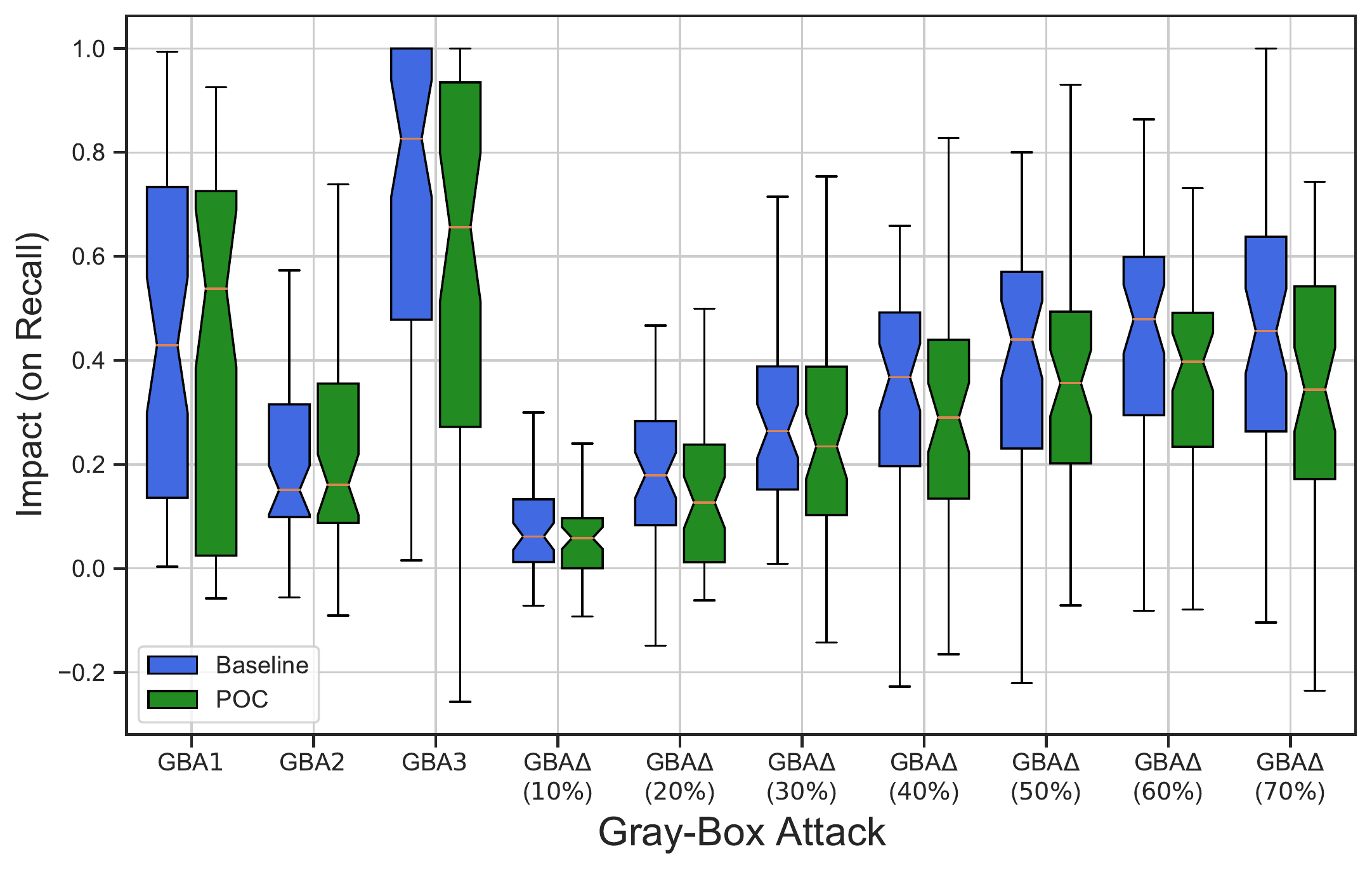}
    \caption{Aggregated \textit{Impact} on all the baseline and \POCscript\ classifiers.}
    \label{fig:boxplot}
\end{figure}

Finally, Figure~\ref{fig:boxplot} shows the aggregated results of all our attacks on all datasets and classifiers. Specifically, Figure~\ref{fig:boxplot} shows 10 pairs of boxplots: each pair represents one of our considered attacks (the 3 simple, and the 7 complex attacks). The blue (resp. green) boxplot of each pair represents the distribution of the \textit{Impact} of the corresponding attack against the baseline (resp. \POC) classifiers (we exclude the few outliers). The figure shows that, in general, the \POC\ classifiers are less affected by the attacks.

All tables showing the \textit{Impact} of the attacks (i.e., Tables~\ref{tab:gba-impact} to~\ref{tab:gba4-difference}) are provided with the same format as those of the \textit{no attack} case (Table~\ref{tab:expt-baseline2} and Table~\ref{tab:expt-trans2}) in the supplementary material.

\subsection{Run Time of Training Phase}
\label{ssec:time}

Table~\ref{tab:time} shows the time (in seconds) required to train the baseline version and the \POC-hardened variant of each classifier.\footnote{Experiments were performed on an Intel 7700HQ CPU (4 cores, 8 threads, 2.8 GHz) with 32GB RAM. We parallelized computations of those classifiers that support multiprocessing (according to scikit-learn).}

\textbf{Overview.} We see that neural net classifiers (MLP, DnW) require the most time to train, regardless of whether \POC\ is applied or not. 
As these classifiers do not provide great detection performance (as shown in previous sections), we do not recommend them as phishing detectors.

\input{sections/tables_big/runtime_table}

\textbf{Baseline vs \POC.}
Surprisingly, the training time of \POC\ is comparable to that of the corresponding baseline. \POC\ requires slightly more time on \LNU\ and UCI, but  slightly less time on Mendeley and DeltaPhish. We reiterate that \POC\ must be trained. If random hyper-parameters are used, it is unlikely to yield great performance (i.e., low false positives and high true positives). The results in Table~\ref{tab:time} denote the time required to train the `best' configuration of \POC\ after our extensive grid-search optimization. Real-world deployments must train. Fortunately, training only needs to be done infrequently (e.g. when hackers change their phishing methods and retraining is needed to prevent concept-drift~\cite{jordaney2017transcend}).

%% file: sections/tables_big/noAtk_tables.tex
\begin{table*}
    \centering
    \caption{No Attack Case: Results of the `baseline' PDs for each dataset and classifier (using all features in the dataset).} 
    
    \resizebox{2\columnwidth}{!}{
        \begin{tabular}{c|| c|c|c|c ? c|c|c|c ? c|c|c|c ? c|c|c|c|}
            \cline{2-17}
            \multicolumn{1}{c}{} & 
            \multicolumn{4}{|c?}{\textbf{\LNUscript}} &
            \multicolumn{4}{|c?}{\textbf{DeltaPhish}} &
            \multicolumn{4}{|c?}{\begin{tabular}{@{}c@{}} \textbf{Mendeley} \\ \textbf{Phishing}\end{tabular}} & 
            \multicolumn{4}{|c|}{\begin{tabular}{@{}c@{}} \textbf{UCI} \\ \textbf{Phishing}\end{tabular}}  \\
            \hline
            Classifier &
            \textit{F1-score} & \textit{Acc} & \textit{FPR} & \textit{TPR} &
            \textit{F1-score} & \textit{Acc} & \textit{FPR} & \textit{TPR} &
            \textit{F1-score} & \textit{Acc} & \textit{FPR} & \textit{TPR} &
            \textit{F1-score} & \textit{Acc} & \textit{FPR} & \textit{TPR} \\
            \midrule

            RF & \small{$0.973$} & \small{$0.982$} & \small{$0.013$} & \small{$0.972$} & \small{$0.959$} & \small{$0.989$} & \small{$0.001$} & \small{$0.926$} & \small{$0.978$} & \small{$0.975$} & \small{$0.033$} & \small{$0.989$} & \small{$0.973$} & \small{$0.973$} & \small{$0.045$} & \small{$0.975$} \\
            SVM & \small{$0.983$} & \small{$0.989$} & \small{$0.004$} & \small{$0.974$} & \small{$0.450$} & \small{$0.876$} & \small{$0.025$} & \small{$0.319$} & \small{$0.927$} & \small{$0.928$} & \small{$0.092$} & \small{$0.943$} & \small{$0.943$} & \small{$0.936$} & \small{$0.113$} & \small{$0.958$} \\
            KNN & \small{$0.996$} & \small{$0.998$} & \small{$0.002$} & \small{$0.997$} & \small{$0.971$} & \small{$0.991$} & \small{$0.007$} & \small{$0.971$} & \small{$0.974$} & \small{$0.975$} & \small{$0.037$} & \small{$0.984$} & \small{$0.984$} & \small{$0.983$} & \small{$0.045$} & \small{$0.997$} \\
            SGD & \small{$0.985$} & \small{$0.990$} & \small{$0.003$} & \small{$0.977$} & \small{$0.458$} & \small{$0.874$} & \small{$0.030$} & \small{$0.333$} & \small{$0.931$} & \small{$0.932$} & \small{$0.083$} & \small{$0.943$} & \small{$0.939$} & \small{$0.932$} & \small{$0.114$} & \small{$0.951$} \\
            DT & \small{$0.986$} & \small{$0.991$} & \small{$0.006$} & \small{$0.985$} & \small{$0.988$} & \small{$0.996$} & \small{$0.002$} & \small{$0.985$} & \small{$0.946$} & \small{$0.948$} & \small{$0.056$} & \small{$0.948$} & \small{$0.990$} & \small{$0.989$} & \small{$0.017$} & \small{$0.991$} \\
            LR & \small{$0.985$} & \small{$0.990$} & \small{$0.003$} & \small{$0.977$} & \small{$0.521$} & \small{$0.776$} & \small{$0.293$} & \small{$0.765$} & \small{$0.937$} & \small{$0.938$} & \small{$0.073$} & \small{$0.946$} & \small{$0.942$} & \small{$0.935$} & \small{$0.115$} & \small{$0.957$} \\
            NB & \small{$0.955$} & \small{$0.971$} & \small{$0.010$} & \small{$0.932$} & \small{$0.714$} & \small{$0.915$} & \small{$0.050$} & \small{$0.667$} & \small{$0.802$} & \small{$0.832$} & \small{$0.046$} & \small{$0.700$} & \small{$0.651$} & \small{$0.715$} & \small{$0.008$} & \small{$0.485$} \\
            MLP & \small{$0.990$} & \small{$0.994$} & \small{$0.001$} & \small{$0.983$} & \small{$0.929$} & \small{$0.978$} & \small{$0.007$} & \small{$0.892$} & \small{$0.976$} & \small{$0.976$} & \small{$0.023$} & \small{$0.975$} & \small{$0.984$} & \small{$0.983$} & \small{$0.026$} & \small{$0.985$} \\
            AB & \small{$0.986$} & \small{$0.991$} & \small{$0.002$} & \small{$0.977$} & \small{$0.872$} & \small{$0.961$} & \small{$0.019$} & \small{$0.833$} & \small{$0.949$} & \small{$0.951$} & \small{$0.054$} & \small{$0.952$} & \small{$0.947$} & \small{$0.941$} & \small{$0.107$} & \small{$0.962$} \\
            ET & \small{$0.999$} & \small{$0.999$} & \small{$0.001$} & \small{$0.999$} & \small{$0.988$} & \small{$0.996$} & \small{$0.001$} & \small{$0.980$} & \small{$0.991$} & \small{$0.991$} & \small{$0.006$} & \small{$0.988$} & \small{$0.991$} & \small{$0.990$} & \small{$0.015$} & \small{$0.992$} \\
            GB & \small{$0.999$} & \small{$0.999$} & \small{$0.001$} & \small{$0.999$} & \small{$0.995$} & \small{$0.998$} & \small{$0.001$} & \small{$0.990$} & \small{$0.990$} & \small{$0.991$} & \small{$0.009$} & \small{$0.989$} & \small{$0.990$} & \small{$0.988$} & \small{$0.020$} & \small{$0.993$} \\
            DnW & \small{$0.988$} & \small{$0.992$} & \small{$0.002$} & \small{$0.980$} & \small{$0.929$} & \small{$0.980$} & \small{$0.005$} & \small{$0.886$} & \small{$0.969$} & \small{$0.970$} & \small{$0.027$} & \small{$0.965$} & \small{$0.974$} & \small{$0.970$} & \small{$0.051$} & \small{$0.981$} \\
            Bag & \small{$0.987$} & \small{$0.992$} & \small{$0.003$} & \small{$0.980$} & \small{$0.983$} & \small{$0.995$} & \small{$0.003$} & \small{$0.980$} & \small{$0.986$} & \small{$0.987$} & \small{$0.012$} & \small{$0.984$} & \small{$0.989$} & \small{$0.988$} & \small{$0.020$} & \small{$0.991$} \\
            \hline
            best & \small{$0.999$} & \small{$0.999$} & \small{$0.001$} & \small{$0.999$} & \small{$0.995$} & \small{$0.998$} & \small{$0.001$} & \small{$0.990$} & \small{$0.991$} & \small{$0.991$} & \small{$0.007$} & \small{$0.989$} & \small{$0.991$} & \small{$0.990$} & \small{$0.023$} & \small{$0.997$} \\

            \bottomrule
            \end{tabular}
            
    }
    \label{tab:expt-baseline2}
\end{table*}

\begin{table*}
    \centering
    \caption{No Attack Case: Results of the \POCscript-hardened PDs for each dataset (using all features in the dataset).}
    
    \resizebox{2\columnwidth}{!}{
        \begin{tabular}{c|| c|c|c|c ? c|c|c|c ? c|c|c|c ? c|c|c|c|}
            \cline{2-17}
            \multicolumn{1}{c}{} & 
            \multicolumn{4}{|c?}{\textbf{\LNUscript}} &
            \multicolumn{4}{|c?}{\textbf{DeltaPhish}} &
            \multicolumn{4}{|c?}{\begin{tabular}{@{}c@{}} \textbf{Mendeley} \\ \textbf{Phishing}\end{tabular}} & 
            \multicolumn{4}{|c|}{\begin{tabular}{@{}c@{}} \textbf{UCI} \\ \textbf{Phishing}\end{tabular}}  \\
            \hline
            Classifier &
            \textit{F1-score} & \textit{Acc} & \textit{FPR} & \textit{TPR} &
            \textit{F1-score} & \textit{Acc} & \textit{FPR} & \textit{TPR} &
            \textit{F1-score} & \textit{Acc} & \textit{FPR} & \textit{TPR} &
            \textit{F1-score} & \textit{Acc} & \textit{FPR} & \textit{TPR} \\
            \midrule
            RF & \small{$0.997$} & \small{$0.998$} & \small{$0.001$} & \small{$0.997$} & \small{$0.990$} & \small{$0.997$} & \small{$0.001$} & \small{$0.980$} & \small{$0.965$} & \small{$0.963$} & \small{$0.033$} & \small{$0.963$} & \small{$0.988$} & \small{$0.987$} & \small{$0.028$} & \small{$0.994$} \\
            SVM & \small{$0.957$} & \small{$0.972$} & \small{$0.009$} & \small{$0.935$} & \small{$0.306$} & \small{$0.866$} & \small{$0.020$} & \small{$0.195$} & \small{$0.682$} & \small{$0.646$} & \small{$0.382$} & \small{$0.715$} & \small{$0.898$} & \small{$0.888$} & \small{$0.184$} & \small{$0.913$} \\
            KNN & \small{$0.997$} & \small{$0.998$} & \small{$0.001$} & \small{$0.997$} & \small{$0.985$} & \small{$0.995$} & \small{$0.004$} & \small{$0.985$} & \small{$0.933$} & \small{$0.925$} & \small{$0.124$} & \small{$0.983$} & \small{$0.986$} & \small{$0.985$} & \small{$0.032$} & \small{$0.993$} \\
            SGD & \small{$0.842$} & \small{$0.888$} & \small{$0.121$} & \small{$0.904$} & \small{$0.409$} & \small{$0.841$} & \small{$0.105$} & \small{$0.365$} & \small{$0.815$} & \small{$0.800$} & \small{$0.208$} & \small{$0.831$} & \small{$0.902$} & \small{$0.892$} & \small{$0.174$} & \small{$0.915$} \\
            DT & \small{$0.989$} & \small{$0.993$} & \small{$0.004$} & \small{$0.986$} & \small{$0.990$} & \small{$0.997$} & \small{$0.001$} & \small{$0.980$} & \small{$0.881$} & \small{$0.875$} & \small{$0.104$} & \small{$0.869$} & \small{$0.987$} & \small{$0.986$} & \small{$0.026$} & \small{$0.991$} \\
            LR & \small{$0.942$} & \small{$0.963$} & \small{$0.008$} & \small{$0.904$} & \small{$0.382$} & \small{$0.689$} & \small{$0.422$} & \small{$0.635$} & \small{$0.781$} & \small{$0.785$} & \small{$0.128$} & \small{$0.723$} & \small{$0.895$} & \small{$0.884$} & \small{$0.187$} & \small{$0.909$} \\
            NB & \small{$0.882$} & \small{$0.921$} & \small{$0.071$} & \small{$0.901$} & \small{$0.723$} & \small{$0.902$} & \small{$0.125$} & \small{$0.850$} & \small{$0.734$} & \small{$0.642$} & \small{$0.599$} & \small{$0.927$} & \small{$0.891$} & \small{$0.879$} & \small{$0.209$} & \small{$0.913$} \\
            MLP & \small{$0.990$} & \small{$0.993$} & \small{$0.001$} & \small{$0.982$} & \small{$0.940$} & \small{$0.982$} & \small{$0.014$} & \small{$0.935$} & \small{$0.883$} & \small{$0.868$} & \small{$0.180$} & \small{$0.933$} & \small{$0.980$} & \small{$0.979$} & \small{$0.045$} & \small{$0.989$} \\
            AB & \small{$0.987$} & \small{$0.991$} & \small{$0.005$} & \small{$0.984$} & \small{$0.921$} & \small{$0.977$} & \small{$0.014$} & \small{$0.900$} & \small{$0.870$} & \small{$0.864$} & \small{$0.108$} & \small{$0.853$} & \small{$0.923$} & \small{$0.915$} & \small{$0.148$} & \small{$0.940$} \\
            ET & \small{$0.998$} & \small{$0.998$} & \small{$0.001$} & \small{$0.997$} & \small{$0.987$} & \small{$0.996$} & \small{$0.001$} & \small{$0.975$} & \small{$0.962$} & \small{$0.960$} & \small{$0.029$} & \small{$0.954$} & \small{$0.988$} & \small{$0.987$} & \small{$0.028$} & \small{$0.994$} \\
            GB & \small{$0.998$} & \small{$0.998$} & \small{$0.001$} & \small{$0.998$} & \small{$0.985$} & \small{$0.995$} & \small{$0.001$} & \small{$0.975$} & \small{$0.961$} & \small{$0.958$} & \small{$0.040$} & \small{$0.962$} & \small{$0.991$} & \small{$0.990$} & \small{$0.022$} & \small{$0.996$} \\
            DnW & \small{$0.987$} & \small{$0.992$} & \small{$0.004$} & \small{$0.983$} & \small{$0.901$} & \small{$0.970$} & \small{$0.006$} & \small{$0.841$} & \small{$0.901$} & \small{$0.912$} & \small{$0.114$} & \small{$0.913$} & \small{$0.968$} & \small{$0.964$} & \small{$0.077$} & \small{$0.985$} \\
            Bag & \small{$0.988$} & \small{$0.992$} & \small{$0.003$} & \small{$0.983$} & \small{$0.990$} & \small{$0.997$} & \small{$0.001$} & \small{$0.980$} & \small{$0.956$} & \small{$0.954$} & \small{$0.044$} & \small{$0.956$} & \small{$0.987$} & \small{$0.986$} & \small{$0.029$} & \small{$0.993$} \\
            \hline
            best & \small{$0.998$} & \small{$0.998$} & \small{$0.001$} & \small{$0.998$} & \small{$0.990$} & \small{$0.997$} & \small{$0.001$} & \small{$0.980$} & \small{$0.962$} & \small{$0.965$} & \small{$0.030$} & \small{$0.963$} & \small{$0.991$} & \small{$0.990$} & \small{$0.022$} & \small{$0.996$} \\
            \bottomrule
        \end{tabular}
    }
    \label{tab:expt-trans2}
\end{table*}

%% file: sections/tables_big/impact-Base_tables.tex
\begin{table*}[!htb]
\centering
    \caption{Simple attack case: \textit{Impact} of \GBA{1} to \GBA{3} on every baseline classifier for each dataset (lower is better).}
    \label{tab:gba-impact}
    \begin{subtable}{0.33\textwidth}
        \centering
        \resizebox{0.9\columnwidth}{!}{
            \begin{tabular}{c|| c ? c ? c ? c |}
                $Clf$ & \textbf{\LNUscript} & \textbf{DeltaPhish} & \textbf{Mendeley} & \textbf{UCI} \\
                \midrule
                RF & $0.124$ & $0.966$ & $0.305$ & $0.750$ \\
                SVM & $0.107$ & $0.784$ & $0.397$ & $0.730$ \\
                KNN & $0.066$ & $0.436$ & $0.017$ & $0.189$ \\
                SGD & $0.121$ & $0.422$ & $0.341$ & $0.730$ \\
                DT & $0.126$ & $0.942$ & $0.813$ & $0.755$ \\
                LR & $0.003$ & $0.080$ & $0.484$ & $0.746$ \\
                NB & $0.267$ & $0.911$ & $0.096$ & $0.506$ \\
                MLP & $0.137$ & $0.696$ & $0.293$ & $0.712$ \\
                AB & $0.114$ & $0.994$ & $0.252$ & $0.748$ \\
                ET & $0.190$ & $0.965$ & $0.619$ & $0.696$ \\
                GB & $0.132$ & $0.984$ & $0.373$ & $0.616$ \\
                DnW & $0.007$ & $0.610$ & $0.249$ & $0.269$ \\
                Bag & $0.124$ & $0.980$ & $0.675$ & $0.723$ \\
                \hline
                average & $0.117$ & $0.751$ & $0.378$ & $0.628$ \\
                \bottomrule
            \end{tabular}
        }
        \vspace{0.1em}
        \caption{\textit{Impact} of \GBA{1}.}
        \label{tab:gba1-impact}
    \end{subtable}%
    \begin{subtable}{0.33\textwidth}
        \centering
        \resizebox{0.9\columnwidth}{!}{
            \begin{tabular}{c|| c ? c ? c ? c |}
                $Clf$ & \textbf{\LNUscript} & \textbf{DeltaPhish} & \textbf{Mendeley} & \textbf{UCI} \\
                \midrule
                RF & $0.290$ & $0.099$ & $0.161$ & $0.112$ \\
                SVM & $0.437$ & $0.103$ & $0.070$ & $0.200$ \\
                KNN & $0.567$ & $0.081$ & $0.970$ & $0.361$ \\
                SGD & $0.672$ & $0.573$ & $0.954$ & $0.157$ \\
                DT & $0.097$ & $0.078$ & $0.121$ & $0.159$ \\
                LR & $0.022$ & $0.270$ & $0.280$ & $0.200$ \\
                NB & $0.364$ & $1.000$ & $-0.056$ & $0.502$ \\
                MLP & $0.048$ & $0.134$ & $0.123$ & $0.272$ \\
                AB & $0.075$ & $0.126$ & $0.128$ & $0.104$ \\
                ET & $0.189$ & $0.139$ & $0.081$ & $0.090$ \\
                GB & $0.172$ & $0.062$ & $0.080$ & $0.455$ \\
                DnW & $0.102$ & $0.004$ & $0.301$ & $0.739$ \\
                Bag & $0.145$ & $0.119$ & $0.732$ & $0.162$ \\
                \hline
                average & $0.244$ & $0.214$ & $0.303$ & $0.270$ \\
                \bottomrule
            \end{tabular}
        }
        \vspace{0.1em}
        \caption{\textit{Impact} of \GBA{2}.}
        \label{tab:gba2-impact}
    \end{subtable}%
    \begin{subtable}{0.33\textwidth}
        \centering
        \resizebox{0.9\columnwidth}{!}{
            \begin{tabular}{c|| c ? c ? c ? c |}
                $Clf$ & \textbf{\LNUscript} & \textbf{DeltaPhish} & \textbf{Mendeley} & \textbf{UCI} \\
                \midrule
                RF & $0.311$ & $1.000$ & $0.669$ & $1.000$ \\
                SVM & $0.540$ & $0.972$ &$0.611$ & $1.000$ \\
                KNN & $0.654$ & $0.709$ &$0.972$ & $0.673$ \\
                SGD & $0.697$ & $0.926$ & $0.998$ & $1.000$ \\
                DT & $0.287$ & $0.942$ & $0.823$ & $1.000$ \\
                LR & $0.015$ & $0.603$ & $0.831$ & $1.000$ \\
                NB & $0.460$ & $1.000$ & $0.062$ & $1.000$ \\
                MLP & $0.171$ & $0.930$ & $0.452$ & $0.998$ \\
                AB & $0.132$ & $1.000$ & $0.481$ & $1.000$ \\
                ET & $0.198$ & $1.000$ & $0.628$ & $1.000$ \\
                GB & $0.185$ & $1.000$ & $0.510$ & $1.000$ \\
                DnW & $0.106$ & $0.652$ & $0.469$ & $0.999$ \\
                Bag & $0.285$ & $1.000$ & $0.951$ & $0.940$ \\
                \hline
                average & $0.310$ & $0.902$ & $0.650$ & $0.97$ \\
                \bottomrule
            \end{tabular}
        }
        \vspace{0.1em}
        \caption{\textit{Impact} of \GBA{3}.}
        \label{tab:gba3-impact}
    \end{subtable} 
\end{table*}

\begin{table*}[htp]
    \centering
    \caption{Complex attack case: \textit{Impact} of the \GBA{$\Delta$} attacks on the baseline PDs for every dataset (lower is better).}
    \label{tab:gba4-impact}
    \begin{subtable}[c]{0.49\textwidth}
        \centering
        \resizebox{\textwidth}{!}{
            \begin{tabular}{c|| c|c|c|c|c|c|c|}
                \cline{2-8}
                \multicolumn{1}{c}{\textbf{\LNUscript}}  & 
                \multicolumn{7}{|c|}{\textbf{Features Modified} ($\Delta$)} \\
                \hline
                Classifier &
                \textit{10\%} & \textit{20\%} & \textit{30\%} &
                \textit{40\%} & \textit{50\%} & \textit{60\%} &
                \textit{70\%} \\
                \midrule
                RF & $0.001$ & $0.044$ & $0.084$ & $0.145$ & $0.173$ & $0.178$ & $0.137$\\
                SVM & $0.072$ & $0.243$ & $0.481$ & $0.496$ & $0.577$ & $0.681$ & $0.686$\\
                KNN & $0.001$ & $0.013$ & $0.022$ & $0.026$ & $0.043$ & $0.043$ & $0.034$\\
                SGD & $0.258$ & $0.278$ & $0.307$ & $0.360$ & $0.425$ & $0.496$ & $0.612$\\
                DT & $-0.002$ & $-0.001$ & $0.099$ & $0.099$ & $0.205$ & $0.204$ & $0.204$\\
                LR & $0.082$ & $0.089$ & $0.239$ & $0.325$ & $0.368$ & $0.387$ & $0.441$\\
                NB & $0.068$ & $0.084$ & $0.125$ & $0.199$ & $0.340$ & $0.519$ & $0.639$\\
                MLP & $0.102$ & $0.113$ & $0.193$ & $0.375$ & $0.450$ & $0.742$ & $0.650$\\
                AB & $0.005$ & $0.005$ & $0.101$ & $0.000$ & $0.000$ & $-0.003$ & $-0.002$\\
                ET & $-0.002$ & $0.005$ & $0.011$ & $0.047$ & $0.090$ & $0.087$ & $0.087$\\
                GB & $0.045$ & $0.049$ & $0.153$ & $0.344$ & $0.369$ & $0.478$ & $0.530$\\
                DnW & $0.012$ & $0.010$ & $0.009$ & $0.056$ & $0.009$ & $0.089$ & $0.264$\\
                Bag & $0.003$ & $0.008$ & $0.219$ & $0.271$ & $0.277$ & $0.277$ & $0.337$\\
                \hline
                average & $0.049$ & $0.072$ & $0.157$ & $0.211$ & $0.255$ & $0.321$ & $0.356$\\
                \bottomrule
            \end{tabular}
        }
        \caption{\GBA{$\Delta$}: \textit{Impact} for the \LNUscript\ dataset.}
        \label{tab:gba4-impact_LNU}
    \end{subtable}
    \hfill
    \begin{subtable}[c]{0.49\textwidth}
        \centering
        \resizebox{\textwidth}{!}{
            \begin{tabular}{c|| c|c|c|c|c|c|c|}
                \cline{2-8}
                \multicolumn{1}{c}{\textbf{DeltaPhish}}  & 
                \multicolumn{7}{|c|}{\textbf{Features Modified ($\Delta$)}} \\
                \hline
                Classifier &
                \textit{10\%} & \textit{20\%} & \textit{30\%} &
                \textit{40\%} & \textit{50\%} & \textit{60\%} &
                \textit{70\%} \\
                \midrule
                RF & $0.038$ & $0.198$ & $0.356$ & $0.389$ & $0.678$ & $0.599$ & $0.567$\\
                SVM & $0.049$ & $0.221$ & $0.037$ & $0.046$ & $-0.123$ & $-0.082$ & $-0.405$\\
                KNN & $0.002$ & $0.157$ & $0.327$ & $0.491$ & $0.568$ & $0.699$ & $0.557$\\
                SGD & $-0.064$ & $0.106$ & $0.267$ & $0.111$ & $0.302$ & $0.484$ & $0.112$\\
                DT & $-0.001$ & $0.080$ & $0.110$ & $0.191$ & $0.189$ & $0.270$ & $0.269$\\
                LR & $0.135$ & $0.327$ & $0.522$ & $0.543$ & $0.625$ & $0.659$ & $0.523$\\
                NB & $0.037$ & $0.105$ & $0.254$ & $0.474$ & $0.631$ & $0.640$ & $0.693$\\
                MLP & $0.081$ & $0.211$ & $0.317$ & $0.389$ & $0.528$ & $0.626$ & $0.763$\\
                AB & $-0.023$ & $0.167$ & $0.224$ & $0.223$ & $0.176$ & $0.599$ & $0.637$\\
                ET & $0.010$ & $0.096$ & $0.235$ & $0.254$ & $0.503$ & $0.583$ & $0.687$\\
                GB & $0.018$ & $0.080$ & $0.138$ & $0.220$ & $0.292$ & $0.376$ & $0.415$\\
                DnW & $0.012$ & $0.010$ & $0.009$ & $0.011$ & $0.009$ & $0.098$ & $0.265$\\
                Bag & $0.087$ & $0.150$ & $0.298$ & $0.319$ & $0.351$ & $0.393$ & $0.400$\\
                \hline
                average & $0.030$ & $0.147$ & $0.238$ & $0.282$ & $0.363$ & $0.458$ & $0.421$\\
                \bottomrule
            \end{tabular}
        }
        \caption{\GBA{$\Delta$}: \textit{Impact} for the DeltaPhish dataset.}
        \label{tab:gba4-impact_DeltaPhish}
    \end{subtable}
\vskip \baselineskip
    \begin{subtable}[c]{0.49\textwidth}
        \centering
        \resizebox{\textwidth}{!}{
            \begin{tabular}{c|| c|c|c|c|c|c|c|}
                \cline{2-8}
                \multicolumn{1}{c}{\textbf{Mendeley}}  & 
                \multicolumn{7}{|c|}{\textbf{Features Modified} ($\Delta$)} \\
                \hline
                Classifier &
                \textit{10\%} & \textit{20\%} & \textit{30\%} &
                \textit{40\%} & \textit{50\%} & \textit{60\%} &
                \textit{70\%} \\
                \midrule
                RF & $0.033$ & $0.235$ & $0.293$ & $0.474$ & $0.618$ & $0.593$ & $0.596$\\
                SVM & $0.189$ & $0.317$ & $0.345$ & $0.299$ & $0.467$ & $0.596$ & $0.643$\\
                KNN & $0.132$ & $0.240$ & $0.408$ & $0.631$ & $0.653$ & $0.706$ & $0.700$\\
                SGD & $0.041$ & $0.212$ & $0.218$ & $0.184$ & $0.245$ & $0.320$ & $0.336$\\
                DT & $0.095$ & $0.247$ & $0.371$ & $0.461$ & $0.517$ & $0.585$ & $0.522$\\
                LR & $0.082$ & $0.117$ & $0.213$ & $0.397$ & $0.433$ & $0.393$ & $0.461$\\
                NB & $-0.072$ & $-0.148$ & $-0.233$ & $-0.227$ & $-0.202$ & $-0.169$ & $-0.105$\\
                MLP & $0.118$ & $0.141$ & $0.148$ & $0.221$ & $0.239$ & $0.300$ & $0.374$\\
                AB & $0.049$ & $0.156$ & $0.226$ & $0.375$ & $0.450$ & $0.548$ & $0.453$\\
                ET & $0.196$ & $0.345$ & $0.564$ & $0.659$ & $0.758$ & $0.672$ & $0.783$\\
                GB & $0.034$ & $0.116$ & $0.261$ & $0.323$ & $0.406$ & $0.431$ & $0.614$\\
                DnW & $0.037$ & $0.081$ & $0.185$ & $0.271$ & $0.456$ & $0.568$ & $0.631$\\
                Bag & $0.129$ & $0.444$ & $0.570$ & $0.580$ & $0.746$ & $0.805$ & $0.886$\\
                \hline
                average & $0.083$ & $0.193$ & $0.275$ & $0.358$ & $0.445$ & $0.488$ & $0.531$\\
                \bottomrule
            \end{tabular}
        }
        \caption{\GBA{$\Delta$}: \textit{Impact} for the Mendeley dataset.}
        \label{tab:gba4-impact_Mendeley}
    \end{subtable}
    \hfill
    \begin{subtable}[c]{0.49\textwidth}
        \centering
        \resizebox{\textwidth}{!}{
            \begin{tabular}{c|| c|c|c|c|c|c|c|}
                \cline{2-8}
                \multicolumn{1}{c}{\textbf{UCI}}  & 
                \multicolumn{7}{|c|}{\textbf{Features Modified} ($\Delta$)} \\
                \hline
                Classifier &
                \textit{10\%} & \textit{20\%} & \textit{30\%} &
                \textit{40\%} & \textit{50\%} & \textit{60\%} &
                \textit{70\%} \\
                \midrule
                RF & $0.022$ & $0.304$ & $0.554$ & $0.481$ & $0.553$ & $0.329$ & $0.038$\\
                SVM & $0.135$ & $0.337$ & $0.501$ & $0.577$ & $0.537$ & $0.352$ & $0.338$\\
                KNN & $0.163$ & $0.374$ & $0.481$ & $0.557$ & $0.800$ & $0.864$ & $0.814$\\
                SGD & $0.201$ & $0.298$ & $0.371$ & $0.532$ & $0.657$ & $0.601$ & $0.400$\\
                DT & $0.300$ & $0.390$ & $0.403$ & $0.389$ & $0.435$ & $0.480$ & $1.000$\\
                LR & $0.199$ & $0.429$ & $0.435$ & $0.604$ & $0.445$ & $0.316$ & $0.220$\\
                NB & $0.400$ & $0.350$ & $0.237$ & $0.085$ & $-0.221$ & $-0.422$ & $-0.735$\\
                MLP & $0.254$ & $0.467$ & $0.715$ & $0.639$ & $0.695$ & $0.331$ & $0.261$\\
                AB & $0.200$ & $0.310$ & $0.401$ & $0.389$ & $0.454$ & $0.312$ & $0.334$\\
                ET & $0.110$ & $0.212$ & $0.348$ & $0.607$ & $0.110$ & $0.085$ & $0.086$\\
                GB & $0.123$ & $0.270$ & $0.364$ & $0.612$ & $0.730$ & $0.565$ & $0.464$\\
                DnW & $0.205$ & $0.250$ & $0.384$ & $0.491$ & $0.472$ & $0.586$ & $0.627$\\
                Bag & $0.056$ & $0.192$ & $0.409$ & $0.517$ & $0.616$ & $0.666$ & $0.695$\\
                \hline
                average & $0.182$ & $0.322$ & $0.431$ & $0.498$ & $0.483$ & $0.390$ & $0.350$\\
                \bottomrule
            \end{tabular}
        }
        \caption{\GBA{$\Delta$}: \textit{Impact} for the UCI dataset.}
        \label{table:gba4-impact_UCI}
    \end{subtable}  
\end{table*}

%% file: sections/tables_big/impact-POC_tables.tex
\begin{table*}[!htb]
\centering
    \caption{Simple attack case. Differences between the \textit{Impact} of \GBA{1} to \GBA{3} on the baselines and on \POCscript\ (higher is better). }
    \label{tab:gba-difference}
    \begin{subtable}{0.33\textwidth}
        \centering
        \resizebox{0.9\columnwidth}{!}{
            \begin{tabular}{c|| c ? c ? c ? c |}
                $Clf$ & \textbf{\LNUscript} & \textbf{DeltaPhish} & \textbf{Mendeley} & \textbf{UCI} \\
                \midrule
                RF & $\bm{0.102}$ & $\bm{0.061}$ & \cellcolor{gray!20}$\bm{0.284}$ & $\bm{0.075}$ \\
                SVM & $\bm{0.115}$ & \cellcolor{gray!45}$\bm{0.791}$ & \cellcolor{gray!20}$\bm{0.372}$ & $\bm{0.080}$ \\
                KNN & $\bm{0.044}$ & $\bm{0.157}$ & $-0.030$ & $\bm{0.013}$ \\
                SGD & $\bm{0.179}$ & $\bm{0.001}$ & \cellcolor{gray!20}$\bm{0.349}$ & $\bm{0.027}$ \\
                DT & $\bm{0.130}$ & $\bm{0.110}$ & \cellcolor{gray!45}$\bm{0.807}$ & $\bm{0.030}$ \\
                LR & $\bm{0.028}$ & $\bm{0.076}$ & \cellcolor{gray!20}$\bm{0.477}$ & $\bm{0.025}$ \\
                NB & $\bm{0.133}$ & \cellcolor{gray!20}$\bm{0.348}$ & $\bm{0.096}$ & $-0.152$ \\
                MLP & $\bm{0.124}$ & \cellcolor{gray!20}$\bm{0.273}$ & \cellcolor{gray!20}$\bm{0.284}$ & $-0.073$ \\
                AB & $\bm{0.089}$ & $\bm{0.068}$ & \cellcolor{gray!20}$\bm{0.252}$ & \cellcolor{gray!20}$\bm{0.235}$ \\
                ET & $\bm{0.171}$ & $\bm{0.115}$ & \cellcolor{gray!45}$\bm{0.544}$ & $\bm{0.086}$\\
                GB & $\bm{0.115}$ & $\bm{0.084}$ & \cellcolor{gray!20}$\bm{0.342}$ & $\bm{0.106}$ \\
                DnW & $-0.001$ & $-0.015$ & $\bm{0.050}$ & $\bm{0.030}$ \\
                Bag & $\bm{0.120}$ & $\bm{0.083}$ & \cellcolor{gray!45}$\bm{0.633}$ & \cellcolor{gray!20}$\bm{0.225}$ \\
                \hline
                average & $\bm{0.103}$ & $\bm{0.166}$ & \cellcolor{gray!20}$\bm{0.344}$ & $\bm{0.055}$ \\
                \bottomrule
            \end{tabular}
        }
        \vspace{0.1em}
        \caption{\GBA{1}: \textit{Impact} difference.}
        \label{tab:gba1-difference}
    \end{subtable}%
    \begin{subtable}{0.33\textwidth}
        \centering
        \resizebox{0.9\columnwidth}{!}{
            \begin{tabular}{c|| c ? c ? c ? c |}
                $Clf$ & \textbf{\LNUscript} & \textbf{DeltaPhish} & \textbf{Mendeley} & \textbf{UCI} \\
                \midrule
                RF & $\bm{0.082}$ & $-0.365$ & $\bm{0.058}$ & $\bm{0.057}$ \\
                SVM & \cellcolor{gray!20}$\bm{0.426}$ & $\bm{0.026}$ & \cellcolor{gray!20}$\bm{0.377}$ & $\bm{0.191}$ \\
                KNN & $\bm{0.048}$ & $-0.876$ & \cellcolor{gray!45}$\bm{0.903}$ & \cellcolor{gray!20}$\bm{0.256}$ \\
                SGD & \cellcolor{gray!45}$\bm{0.763}$ & $-0.223$ & \cellcolor{gray!45}$\bm{0.682}$ & $\bm{0.137}$ \\
                DT & $-0.001$ & $-0.210$ & $\bm{0.032}$ & $\bm{0.009}$ \\
                LR & $\bm{0.101}$ & \cellcolor{gray!20}$\bm{0.274}$ & \cellcolor{gray!20}$\bm{0.290}$ & $\bm{0.197}$ \\
                NB & \cellcolor{gray!20}$\bm{0.270}$ & $\bm{0.077}$ & $-0.125$ & \cellcolor{gray!45}$\bm{0.540}$ \\
                MLP & $-0.106$ & $\bm{0.054}$ & $\bm{0.058}$ & $\bm{0.079}$ \\
                AB & $-0.019$ & $-0.408$ & $\bm{0.067}$ & $\bm{0.116}$ \\
                ET & $-0.011$ & $-0.215$ & $-0.061$ & $\bm{0.034}$ \\
                GB & $-0.002$ & $-0.398$ & $-0.023$ & \cellcolor{gray!20}$\bm{0.223}$ \\
                DnW & $-0.03$ & $-0.219$ & \cellcolor{gray!20}$\bm{0.288}$ & $\bm{0.130}$ \\
                Bag & $-0.059$ & $-0.386$ & \cellcolor{gray!45}$\bm{0.591}$ & $\bm{0.058}$ \\
                \hline
                average & $\bm{0.112}$ & $-0.220$ & \cellcolor{gray!20}$\bm{0.241}$ & $\bm{0.156}$ \\
                \bottomrule
            \end{tabular}
        }
        \vspace{0.1em}
        \caption{\GBA{2}: \textit{Impact} difference.}
        \label{tab:gba2-difference}
    \end{subtable}%
    \begin{subtable}{0.33\textwidth}
        \centering
        \resizebox{0.9\columnwidth}{!}{
            \begin{tabular}{c|| c ? c ? c ? c |}
                $Clf$ & \textbf{\LNUscript} & \textbf{DeltaPhish} & \textbf{Mendeley} & \textbf{UCI} \\
                \midrule
                RF & $-0.005$ & $\bm{0.108}$ & \cellcolor{gray!20}$\bm{0.348}$ & \cellcolor{gray!20}$\bm{0.242}$\\
                SVM & \cellcolor{gray!45}$\bm{0.590}$ & $\bm{0.080}$ & \cellcolor{gray!20}$\bm{0.415}$ & \cellcolor{gray!45}$\bm{0.845}$\\
                KNN & $\bm{0.040}$ & \cellcolor{gray!45}$\bm{0.538}$ & \cellcolor{gray!45}$\bm{0.905}$ & \cellcolor{gray!20}$\bm{0.420}$\\
                SGD & $\bm{0.357}$ & \cellcolor{gray!45}$\bm{0.511}$ & $\bm{0.038}$ & \cellcolor{gray!45}$\bm{0.806}$\\
                DT & $\bm{0.152}$ & \cellcolor{gray!20}$\bm{0.223}$ & \cellcolor{gray!45}$\bm{0.569}$ & \cellcolor{gray!20}$\bm{0.484}$\\
                LR & $-0.43$ & \cellcolor{gray!45}$\bm{0.859}$ & \cellcolor{gray!45}$\bm{0.958}$ & \cellcolor{gray!45}$\bm{0.831}$\\
                NB & $\bm{0.038}$ & $\bm{0.067}$ & $\bm{0.127}$ & \cellcolor{gray!45}$\bm{0.987}$\\
                MLP & $-0.059$ & $\bm{0.129}$ & $\bm{0.092}$ & \cellcolor{gray!45}$\bm{0.508}$\\
                AB & $\bm{0.025}$ & $\bm{0.046}$ & \cellcolor{gray!20}$\bm{0.248}$ & \cellcolor{gray!45}$\bm{0.550}$\\
                ET & $-0.081$ & $\bm{0.185}$ & \cellcolor{gray!20}$\bm{0.284}$ & \cellcolor{gray!20}$\bm{0.450}$\\
                GB & $\bm{0.021}$ & $\bm{0.075}$ & $\bm{0.161}$ & \cellcolor{gray!20}$\bm{0.237}$\\
                DnW & $-0.011$ & $-0.019$ & $\bm{0.155}$ & \cellcolor{gray!20}$\bm{0.356}$ \\
                Bag & $-0.036$ & $\bm{0.116}$ & \cellcolor{gray!45}$\bm{0.625}$ & \cellcolor{gray!20}$\bm{0.233}$\\
                \hline
                average & $\bm{0.046}$ & \cellcolor{gray!20}$\bm{0.224}$ & \cellcolor{gray!20}$\bm{0.378}$ & \cellcolor{gray!45}$\bm{0.535}$\\
                \bottomrule
            \end{tabular}
        }
        \vspace{0.1em}
        \caption{\GBA{3}: \textit{Impact} difference.}
        \label{tab:gba3-difference}
    \end{subtable} 
\end{table*}

\begin{table*}[htp]
    \centering
    \caption{Complex attack case. Differences between the \textit{Impact} of the \GBA{$\Delta$} attack on the baselines and on \POCscript\ (higher is better).}
    \label{tab:gba4-difference}
    \begin{subtable}[c]{0.49\textwidth}
        \centering
        \resizebox{\textwidth}{!}{
            \begin{tabular}{c|| c|c|c|c|c|c|c|}
                \cline{2-8}
                \multicolumn{1}{c}{\textbf{\LNUscript}}  & 
                \multicolumn{7}{|c|}{\textbf{Features Modified} ($\Delta$)} \\
                \hline
                Classifier &
                \textit{10\%} & \textit{20\%} & \textit{30\%} &
                \textit{40\%} & \textit{50\%} & \textit{60\%} &
                \textit{70\%} \\
                \midrule
                RF & $\bm{0.003}$ & $\bm{0.004}$ & $-0.003$ & $-0.111$ & $-0.197$ & $-0.315$ & $-0.408$\\
                SVM & $\bm{0.048}$ & $\bm{0.107}$ & $\bm{0.196}$ & \cellcolor{gray!20}$\bm{0.210}$ & $\bm{0.150}$ & $\bm{0.146}$ & $\bm{0.084}$\\
                KNN & $\bm{0.004}$ & $\bm{0.004}$ & $\bm{0.005}$ & $\bm{0.003}$ & $\bm{0.005}$ & $\bm{0.008}$ & $\bm{0.002}$\\
                SGD & $\bm{0.183}$ & $\bm{0.179}$ & \cellcolor{gray!20}$\bm{0.209}$ & $\bm{0.161}$ & $\bm{0.144}$ & $\bm{0.104}$ & $\bm{0.070}$\\
                DT & $\bm{0.000}$ & $\bm{0.001}$ & $\bm{0.006}$ & $-0.029$ & $-0.034$ & $-0.237$ & $-0.133$\\
                LR & $\bm{0.054}$ & $\bm{0.017}$ & $\bm{0.091}$ & $\bm{0.118}$ & $\bm{0.096}$ & $\bm{0.062}$ & $\bm{0.069}$\\
                NB & $\bm{0.067}$ & $\bm{0.076}$ & $\bm{0.098}$ & $\bm{0.116}$ & $\bm{0.115}$ & $\bm{0.056}$ & $\bm{0.063}$\\
                MLP & $\bm{0.001}$ & $\bm{0.105}$ & $\bm{0.093}$ & $\bm{0.090}$ & \cellcolor{gray!20}$\bm{0.247}$ & \cellcolor{gray!20}$\bm{0.499}$ & \cellcolor{gray!45}$\bm{0.559}$\\
                AB & $\bm{0.008}$ & $\bm{0.008}$ & $-0.027$ & $-0.136$ & $-0.166$ & $-0.370$ & $-0.368$\\
                ET & $\bm{0.002}$ & $-0.001$ & $0.000$ & $\bm{0.005}$ & $\bm{0.006}$ & $-0.075$ & $-0.188$\\
                GB & $\bm{0.013}$ & $\bm{0.013}$ & $\bm{0.011}$ & $\bm{0.107}$ & $\bm{0.053}$ & $\bm{0.171}$ & $\bm{0.189}$\\
                DnW & $-0.015$ & $-0.004$ & $-0.012$ & $\bm{0.03}$ & $-0.004$ & $\bm{0.005}$ & $\bm{0.025}$\\
                Bag & $\bm{0.003}$ & $\bm{0.003}$ & $\bm{0.014}$ & $\bm{0.104}$ & \cellcolor{gray!20}$\bm{0.209}$ & \cellcolor{gray!20}$\bm{0.209}$ & \cellcolor{gray!20}$\bm{0.305}$\\
                \hline
                average & $\bm{0.029}$ & $\bm{0.039}$ & $\bm{0.053}$ & $\bm{0.051}$ & $\bm{0.048}$ & $\bm{0.021}$ & $\bm{0.021}$\\
                \bottomrule
            \end{tabular}
        }
        \caption{\GBA{$\Delta$}: \textit{Impact} differences for the \LNUscript~dataset.}
        \label{tab:gba4-difference_LNU}
        \vspace{1.5mm}
    \end{subtable}
    \hfill
    \begin{subtable}[c]{0.49\textwidth}
        \centering
        \resizebox{\textwidth}{!}{
            \begin{tabular}{c|| c|c|c|c|c|c|c|}
                \cline{2-8}
                \multicolumn{1}{c}{\textbf{DeltaPhish}}  & 
                \multicolumn{7}{|c|}{\textbf{Features Modified} ($\Delta$)} \\
                \hline
                Classifier &
                \textit{10\%} & \textit{20\%} & \textit{30\%} &
                \textit{40\%} & \textit{50\%} & \textit{60\%} &
                \textit{70\%} \\
                \midrule
                RF & $-0.032$ & $\bm{0.072}$ & $\bm{0.161}$ & $\bm{0.127}$ & \cellcolor{gray!20}$\bm{0.260}$ & $\bm{0.157}$ & $\bm{0.054}$\\
                SVM & \cellcolor{gray!20}$\bm{0.311}$ & \cellcolor{gray!20}$\bm{0.283}$ & \cellcolor{gray!20}$\bm{0.425}$ & \cellcolor{gray!45}$\bm{0.532}$ & \cellcolor{gray!70}$\bm{0.753}$ & \cellcolor{gray!70}$\bm{0.764}$ & $\bm{0.251}$\\
                KNN & $\bm{0.024}$ & $\bm{0.169}$ & $\bm{0.197}$ & \cellcolor{gray!20}$\bm{0.369}$ & \cellcolor{gray!20}$\bm{0.391}$ & \cellcolor{gray!20}$\bm{0.296}$ & \cellcolor{gray!20}$\bm{0.314}$\\
                SGD & $\bm{0.188}$ & $\bm{0.164}$ & \cellcolor{gray!20}$\bm{0.409}$ & $\bm{0.027}$ & \cellcolor{gray!20}$\bm{0.374}$ & \cellcolor{gray!20}$\bm{0.291}$ & \cellcolor{gray!20}$\bm{0.347}$\\
                DT & $-0.084$ & $-0.141$ & $-0.238$ & $-0.211$ & $-0.220$ & $-0.231$ & $-0.219$\\
                LR & $\bm{0.041}$ & $\bm{0.070}$ & $\bm{0.191}$ & $\bm{0.086}$ & $\bm{0.173}$ & $\bm{0.127}$ & \cellcolor{gray!20}$\bm{0.243}$\\
                NB & $-0.058$ & $-0.067$ & $-0.027$ & $\bm{0.114}$ & $\bm{0.131}$ & $\bm{0.154}$ & $-0.051$\\
                MLP & $-0.130$ & $-0.174$ & $-0.188$ & $-0.140$ & $\bm{0.022}$ & \cellcolor{gray!20}$\bm{0.219}$ & \cellcolor{gray!45}$\bm{0.602}$\\
                AB & $-0.023$ & $-0.022$ & $-0.080$ & $-0.216$ & $-0.405$ & $-0.093$ & $-0.027$\\
                ET & $-0.051$ & $-0.045$ & $-0.044$ & $-0.094$ & $\bm{0.011}$ & $-0.037$ & $\bm{0.033}$\\
                GB & $-0.014$ & $-0.019$ & $-0.077$ & $-0.074$ & $-0.039$ & $-0.042$ & $-0.026$\\
                DnW & $-0.058$ & $-0.003$ & $\bm{0.0310}$ & $\bm{0.037}$ & $-0.010$ & $\bm{0.032}$ & $\bm{0.037}$\\
                Bag & $\bm{0.050}$ & $\bm{0.050}$ & $\bm{0.104}$ & $\bm{0.096}$ & $\bm{0.136}$ & $\bm{0.157}$ & $\bm{0.090}$\\
                \hline
                average & $\bm{0.012}$ & $\bm{0.0259}$ & $\bm{0.066}$ & $\bm{0.051}$ & \cellcolor{gray!20}$\bm{0.121}$ & \cellcolor{gray!20}$\bm{0.138}$ & \cellcolor{gray!20}$\bm{0.126}$\\
                \bottomrule
            \end{tabular}
        }
        \caption{\GBA{$\Delta$}: \textit{Impact} differences for the DeltaPhish dataset.}
        \label{tab:gba4-difference_DeltaPhish}
    \end{subtable}
\vskip\baselineskip
    \begin{subtable}[c]{0.49\textwidth}
        \centering
        \resizebox{\textwidth}{!}{
            \begin{tabular}{c|| c|c|c|c|c|c|c|}
                \cline{2-8}
                \multicolumn{1}{c}{\textbf{Mendeley}}  & 
                \multicolumn{7}{|c|}{\textbf{Features Modified} ($\Delta$)} \\
                \hline
                Classifier &
                \textit{10\%} & \textit{20\%} & \textit{30\%} &
                \textit{40\%} & \textit{50\%} & \textit{60\%} &
                \textit{70\%} \\
                \midrule
                RF & $-0.034$ & $-0.003$ & $\bm{0.011}$ & $\bm{0.099}$ & $\bm{0.113}$ & $-0.002$ & $-0.052$\\
                SVM & \cellcolor{gray!20}$\bm{0.282}$ & \cellcolor{gray!20}$\bm{0.258}$ & $\bm{0.142}$ & \cellcolor{gray!20}$\bm{0.464}$ & $-0.158$ & $\bm{0.040}$ & $\bm{0.200}$\\
                KNN & $\bm{0.078}$ & $\bm{0.112}$ & $\bm{0.150}$ & \cellcolor{gray!20}$\bm{0.292}$ & \cellcolor{gray!20}$\bm{0.264}$ & \cellcolor{gray!20}$\bm{0.288}$ & $-0.037$\\
                SGD & $\bm{0.052}$ & $\bm{0.175}$ & $\bm{0.114}$ & $\bm{0.089}$ & $\bm{0.075}$ & $\bm{0.128}$ & \cellcolor{gray!20}$\bm{0.265}$\\
                DT & $-0.063$ & $\bm{0.027}$ & $-0.086$ & $-0.016$ & $\bm{0.084}$ & $\bm{0.129}$ & $\bm{0.174}$\\
                LR & $\bm{0.097}$ & $\bm{0.126}$ & $\bm{0.127}$ & \cellcolor{gray!20}$\bm{0.234}$ & \cellcolor{gray!20}$\bm{0.232}$ & $\bm{0.125}$ & $-0.106$\\
                NB & $-0.073$ & $-0.135$ & $-0.196$ & $-0.183$ & $-0.143$ & $-0.106$ & $-0.037$\\
                MLP & $-0.008$ & $-0.028$ & $-0.114$ & $-0.182$ & $-0.003$ & $\bm{0.116}$ & $-0.056$\\
                AB & $\bm{0.023}$ & $\bm{0.049}$ & $\bm{0.114}$ & $\bm{0.067}$ & $\bm{0.066}$ & \cellcolor{gray!20}$\bm{0.319}$ & $\bm{0.162}$\\
                ET & $\bm{0.159}$ & $\bm{0.073}$ & $\bm{0.158}$ & \cellcolor{gray!20}$\bm{0.245}$ & \cellcolor{gray!20}$\bm{0.242}$ & $\bm{0.194}$ & $\bm{0.138}$\\
                GB & $-0.044$ & $\bm{0.059}$ & $\bm{0.007}$ & $\bm{0.037}$ & $\bm{0.058}$ & $-0.060$ & $\bm{0.076}$\\
                DnW & $\bm{0.068}$ & $\bm{0.073}$ & $\bm{0.055}$ & $\bm{0.001}$ & $\bm{0.124}$ & $\bm{0.111}$ & $\bm{0.049}$\\
                Bag & $-0.034$ & $\bm{0.173}$ & $\bm{0.155}$ & $\bm{0.062}$ & $\bm{0.115}$ & $\bm{0.096}$ & \cellcolor{gray!20}$\bm{0.242}$\\
                \hline
                average & $\bm{0.039}$ & $\bm{0.074}$ & $\bm{0.049}$ & $\bm{0.093}$ & $\bm{0.083}$ & $\bm{0.106}$ & $\bm{0.079}$\\
                \bottomrule
            \end{tabular}
        }
        \caption{\GBA{$\Delta$}: \textit{Impact} differences for the Mendeley dataset.}
        \label{tab:gba4-difference_Mendeley}
        \vspace{1.5mm}
    \end{subtable}
    \hfill
    \begin{subtable}[c]{0.49\textwidth}
        \centering
        \resizebox{\textwidth}{!}{
            \begin{tabular}{c|| c|c|c|c|c|c|c|}
                \cline{2-8}
                \multicolumn{1}{c}{\textbf{UCI}}  & 
                \multicolumn{7}{|c|}{\textbf{Features Modified} ($\Delta$)} \\
                \hline
                Classifier &
                \textit{10\%} & \textit{20\%} & \textit{30\%} &
                \textit{40\%} & \textit{50\%} & \textit{60\%} &
                \textit{70\%} \\
                \midrule
                RF & $-0.034$ & $\bm{0.002}$ & $\bm{0.140}$ & $-0.199$ & $\bm{0.102}$ & $-0.019$ & $-0.067$\\
                SVM & $-0.070$ & $\bm{0.040}$ & $\bm{0.068}$ & $-0.016$ & $\bm{0.113}$ & $\bm{0.009}$ & $\bm{0.105}$\\
                KNN & $\bm{0.072}$ & $\bm{0.134}$ & $\bm{0.096}$ & $\bm{0.018}$ & $\bm{0.062}$ & $\bm{0.133}$ & \cellcolor{gray!20}$\bm{0.308}$\\
                SGD & $-0.161$ & $-0.176$ & $-0.027$ & \cellcolor{gray!20}$\bm{0.221}$ & \cellcolor{gray!20}$\bm{0.430}$ & \cellcolor{gray!20}$\bm{0.327}$ & \cellcolor{gray!20}$\bm{0.305}$\\
                DT & $\bm{0.060}$ & $-0.007$ & $-0.078$ & $-0.172$ & $-0.063$ & $\bm{0.100}$ & \cellcolor{gray!45}$\bm{0.795}$\\
                LR & $\bm{0.116}$ & $\bm{0.093}$ & $-0.021$ & $\bm{0.015}$ & $-0.115$ & $\bm{0.001}$ & $\bm{0.174}$\\
                NB & $\bm{0.122}$ & $-0.141$ & $-0.220$ & $-0.264$ & $-0.586$ & $-0.343$ & $-0.652$\\
                MLP & $\bm{0.010}$ & $-0.032$ & $-0.039$ & $-0.189$ & $-0.235$ & $-0.568$ & $-0.097$\\
                AB & $-0.007$ & $\bm{0.113}$ & $\bm{0.028}$ & $\bm{0.004}$ & $\bm{0.137}$ & $-0.035$ & $\bm{0.158}$\\
                ET & $\bm{0.026}$ & $\bm{0.008}$ & $\bm{0.173}$ & \cellcolor{gray!45}$\bm{0.553}$ & $\bm{0.045}$ & $\bm{0.048}$ & $-0.136$\\
                GB & $-0.017$ & $\bm{0.016}$ & $\bm{0.006}$ & $\bm{0.010}$ & $\bm{0.174}$ & $\bm{0.115}$ & \cellcolor{gray!20}$\bm{0.401}$\\
                DnW & $\bm{0.015}$ & $\bm{0.052}$ & $-0.014$ & $\bm{0.048}$ & $\bm{0.016}$ & $\bm{0.043}$ & $\bm{0.020}$\\
                Bag & $-0.017$ & $-0.015$ & $-0.035$ & $\bm{0.004}$ & $-0.009$ & $\bm{0.101}$ & \cellcolor{gray!20}$\bm{0.294}$\\
                \hline
                average & $\bm{0.010}$ & $\bm{0.007}$ & $\bm{0.006}$ & $\bm{0.003}$ & $\bm{0.006}$ & $-0.005$ & $\bm{0.124}$\\
                \bottomrule
            \end{tabular}
        }
        \caption{\GBA{$\Delta$}: \textit{Impact} differences for the UCI dataset.}
        \label{tab:gba4-difference_UCI}
    \end{subtable}  
\end{table*}

%% file: sections/tables_big/runtime_table.tex
\begin{table}
    \centering
    \caption{Training Times. For each dataset and PD, we report the time (in seconds) required to train its baseline and \POCscript\ variants.} 
    \label{tab:time}
    \resizebox{\columnwidth}{!}{
        \begin{tabular}{c || c | c ? c | c ? c | c ? c | c}
        \toprule
        Dataset & \multicolumn{2}{c?}{{\LNUscript}} & \multicolumn{2}{c?}{DeltaPhish} & \multicolumn{2}{c?}{Mendeley} & \multicolumn{2}{c?}{UCI} \\ \hline 
        Classifier & Base & \POCscript & Base & \POCscript & Base & \POCscript & Base & \POCscript \\
        \midrule
        RF & \res{0.46} & \res{1.03} & \res{0.50} & \res{0.25} & \res{0.59} & \res{0.24} & \res{0.41} & \res{0.36} \\
        SVM & \res{1.15} & \res{2.66} & \res{0.10} & \res{0.11} & \res{0.75} & \res{0.64} & \res{0.52} & \res{0.77} \\
        KNN & \res{0.06} & \res{0.03} & \res{0.02} & \res{0.01} & \res{0.01} & \res{0.01} & \res{0.11} & \res{0.06} \\
        SGD & \res{0.22} & \res{0.19} & \res{0.01} & \res{0.02} & \res{0.15} & \res{0.06} & \res{0.14} & \res{0.13} \\
        DT & \res{0.07} & \res{0.24} & \res{0.02} & \res{0.04} & \res{0.03} & \res{0.01} & \res{0.02} & \res{0.04} \\
        LR & \res{0.11} & \res{0.30} & \res{0.07} & \res{0.03} & \res{0.89} & \res{0.64} & \res{0.12} & \res{0.07} \\
        NB & \res{0.01} & \res{0.02} & \res{0.01} & \res{0.01} & \res{0.01} & \res{0.01} & \res{0.01} & \res{0.01} \\
        MLP & \res{18.7} & \res{22.7} & \res{8.03} & \res{4.71} & \res{15.7} & \res{13.6} & \res{14.2} & \res{28.9} \\
        AB & \res{0.56} & \res{0.94} & \res{1.24} & \res{1.61} & \res{2.57} & \res{1.44} & \res{0.25} & \res{0.25} \\
        ET & \res{1.32} & \res{1.63} & \res{0.38} & \res{0.38} & \res{0.11} & \res{0.11} & \res{1.13} & \res{1.22} \\
        GB & \res{2.14} & \res{29.7} & \res{2.16} & \res{3.69} & \res{3.33} & \res{4.19} & \res{5.59} & \res{6.79} \\
        DnW & \res{53.4} & \res{62.4} & \res{21.9} & \res{12.7} & \res{22.7} & \res{18.2} & \res{25.0} & \res{35.4} \\
        Bag & \res{0.94} & \res{1.73} & \res{1.24} & \res{0.4} & \res{1.38} & \res{1.31} & \res{1.45} & \res{1.28} \\
        \bottomrule
        
        \end{tabular}
    }
\end{table}

%% file: sections/7-discussion.tex
\section{Discussion}
\label{sec:discussion}

We highlight the key findings from our huge experimental analysis by providing a formal statistical analysis of our results, as well as an in-depth assessment of a pragmatic application of \POC, showing its pros and cons.

\subsection{Statistical Analysis}
\label{ssec:statistical}
We conduct a statistical analysis of our results with the goal of answering three questions:
\begin{enumerate}
    \item is the slight performance drop of \POC\ in the no-attack case significant?
    \item is the \textit{Impact} of the considered attacks on the baseline classifiers significant? 
    \item does \POC\ provide better protection against such attacks than the baseline classifiers?
\end{enumerate}
To answer all these questions, we rely on the Wilcoxon Signed Rank test which performs a pairwise comparison of the samples of two populations. The output of the test is a $p$-value that provides the probability that the two populations were generated by the same underlying process: if the resulting $p$-value is \textit{higher} (resp. lower) than a given target threshold $\alpha$, then the two populations can be considered to be statistically equivalent (resp. different). Typically, $\alpha$ is chosen to be 0.05, meaning the chance of a correct claim is 95\%. 

\subsubsection{No-attack case performance}
We statistically compare the populations of the Recall (i.e., detection rate) and F1-score achieved by the baseline and \POC\ classifiers in the no attack case; all populations consist of 52 elements (given by 13 classifiers and 4 datasets). The resulting $p$-value of these comparisons is 0.37 for the Recall, and 0.23 for the F1-score. Both $p$-values are much higher than $\alpha$, meaning that the populations in both tests can be considered to be statistically equivalent. Therefore, the performance drop of \POC\ is negligible. The reason for this is that our \POC\ implementation in Section~\ref{sec:experiments} assumes complete \textit{prevalence}---meaning that the baseline and \POC\ classifiers use the same amount of information to perform their inference (but the \POC\ variants maps such information in a different space). 

\subsubsection{Impact Assessment} we compare the populations containing the Recall \textit{before} and \textit{after} the execution of each Gray Box attack (hence, the populations have 52 elements for each comparison) on the `baseline' PD. The resulting $p$-value are not only \textit{always} lower than $\alpha$, but are also almost always equal to 0---the only exception is for \GBA{$\Delta$} with $\Delta$=10\%, which has a $p$-value of 0.00003. Therefore, all attacks induce a statistically significant drop in the baseline detection rate. This motivates the search for a solution that mitigates such \textit{Impact}.

\subsubsection{Protection of \POC}
our experiments suggest that hardening classifiers with \POC\ yields results that are superior to those of their baseline variants, but in some cases, the difference is small (see Figure~\ref{fig:boxplot}) and in other cases the baselines are better (i.e., the negative values in Tables~\ref{tab:gba-difference} and~\ref{tab:gba4-difference}). Hence, answering the third question requires a more fine-grained investigation. 
For each dataset, we compare two populations containing the \textit{Impact} of the considered attacks (\GBA{1}-\GBA{3}, and \GBA{$\Delta$} in its 7 variants---10 attacks in all). The first population represents the \textit{Impact} against the baseline classifiers, and the second population represents the \textit{Impact} against the \POC\ versions of the classifiers. Hence, each population has 130 samples (13 $Clf$ * 10 $Att$). Since we distinguish the populations on a per-dataset basis, we apply the Bonferroni Correction, thus resulting in a target $\alpha$=0.0125 (because we are considering 4 different scenarios, one per dataset).
Table~\ref{tab:wilcoxon} shows the resulting $p$-values and Effect~Sizes of the test. We see that all $p$-values are below our target $\alpha$=0.01. Furthermore, the different Effect~Sizes also confirm the low chance that the two populations were generated by the same underlying stochastic process.
The results confirm that using \POC\ yields more resilient classifiers against the Gray Box attacks considered in this paper.

\begin{table}[!htbp]
\centering
\caption{Statistical comparison of the \textit{Impact} against the baselines and \POCscript\ on each dataset (via a Wilcoxon Signed-Rank test).}
\resizebox{\columnwidth}{!}{%
  \begin{tabular}{c||c|c|c|c}
  \toprule
  \textit{Metric} & \LNUscript\ & DeltaPhish & Mendeley & UCI \\
  \midrule
  \textit{p-value} & $<0.0001$ & $0.0005$ & $<0.0001$ & $<0.0001$ \\
  \textit{Effect Size} & $0.2470$ & $0.3256$ & $0.1345$ & $0.2696$ \\
  \bottomrule
  \end{tabular}
  }
  \label{tab:wilcoxon}
\end{table}

Intuitively, \POC\ is effective\footnote{The classifiers are more resilient, but we do not claim that \POCscript\ yields PDs that are immune to such attacks!} against our Gray Box attacks because the baseline PDs use `fixed' features whose modifications result in highly distinct samples. In contrast, when using \POC, the combination of feature mapping and mixing leads to `smoother' feature modifications that do not result in samples deviating greatly from their unmodified variants.
Let us explain this with an example. Suppose a sample $x$ is described by (among others) two binary features $f_1$ and $f_2$, so that $f_1(x)$=$1$ and $f_2(x)$=$1$. Assume an attack that modifies the value of $f_1$ from $1$ to $0$. This translates to an (adversarial) sample $\bar{x}$ whose value of $f_1$ is `the opposite' of its original variant $x$.
Now consider an implementation of \POC\ that, in its $\Psi$, has an $oc$ that sums the two `original features' $f_1$ and $f_2$, implying that its application to $x$ results in $oc(x)$=$2$. If the attacker modifies $f_1$ of sample $x$ from $1$ to $0$, then \textit{after applying} \POC, this modification would result in $oc(\bar{x})$=$1$ which is `less' different from its original variant. To achieve the same effects, the attacker must modify both $f_1$ and $f_2$. This may therefore lead to a more accurate classification.

We note that the `favorable' results in Table~\ref{tab:wilcoxon} are mostly due to the good hardening performance of \POC\ against the simple Gray Box Attacks, i.e., \GBA{1}--\GBA{3} (shown in Table~\ref{tab:gba-difference}). The hardening provided by \POC\ against the complex attacks of \GBA{$\Delta$} (shown in Table~\ref{tab:gba4-difference}) is smaller. Despite this, the next section showcases a pragmatic use-case which shows the low-level benefits of \POC.

\subsection{Pragmatic Use Case}
\label{ssec:pragmatic}

\begin{figure*}[!htbp]
    \centering
    \begin{subfigure}{0.5\columnwidth}
        \centering
        \includegraphics[width=\linewidth]{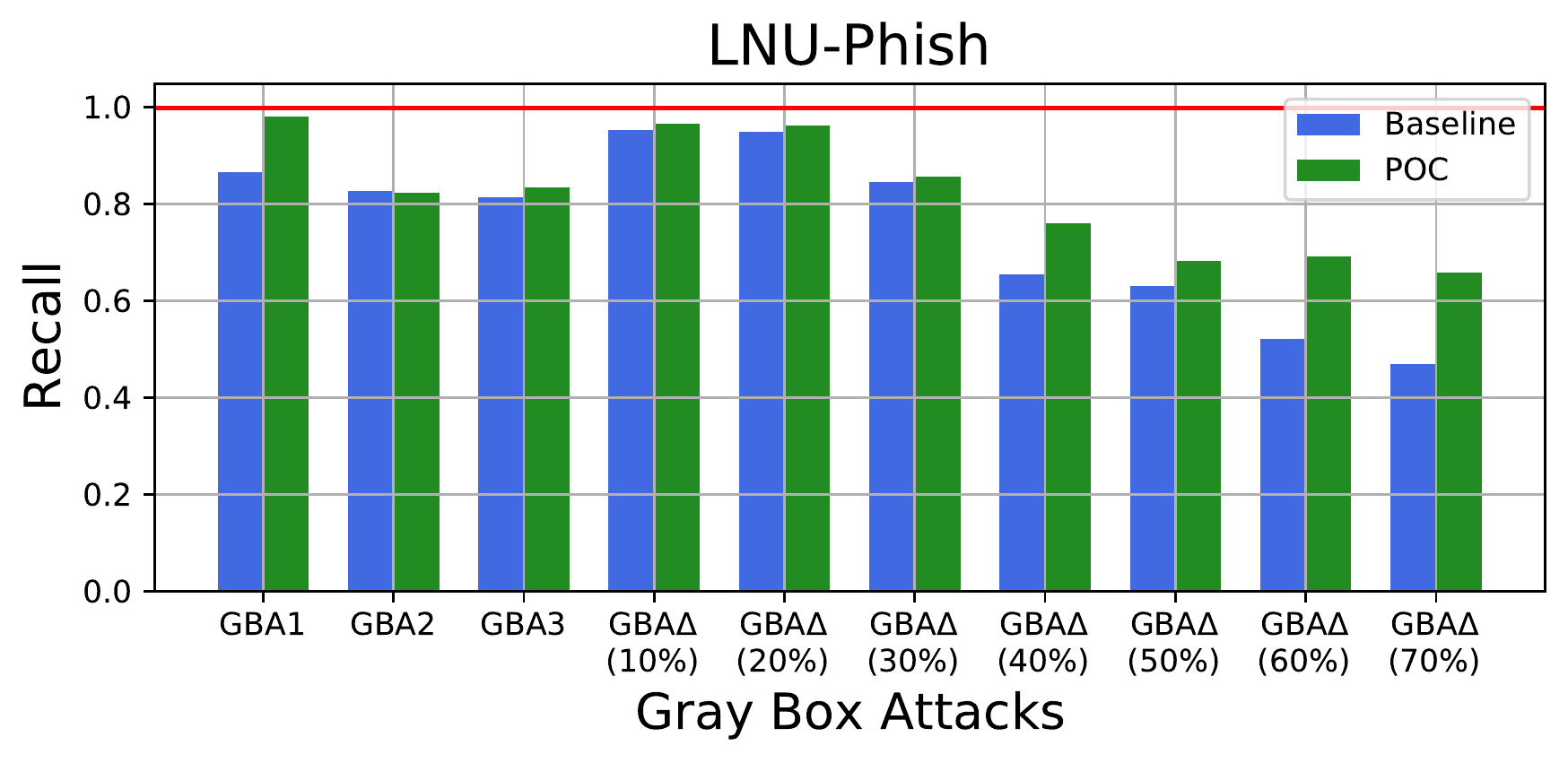}
        \caption{\LNUscript. Best baseline: GB\\(F1-score: 0.99, and 0.99 for \POCscript).}
        \label{sfig:pragmatic_lnu}
    \end{subfigure}\hfill%
    \begin{subfigure}{0.5\columnwidth}
        \centering
        \includegraphics[width=\linewidth]{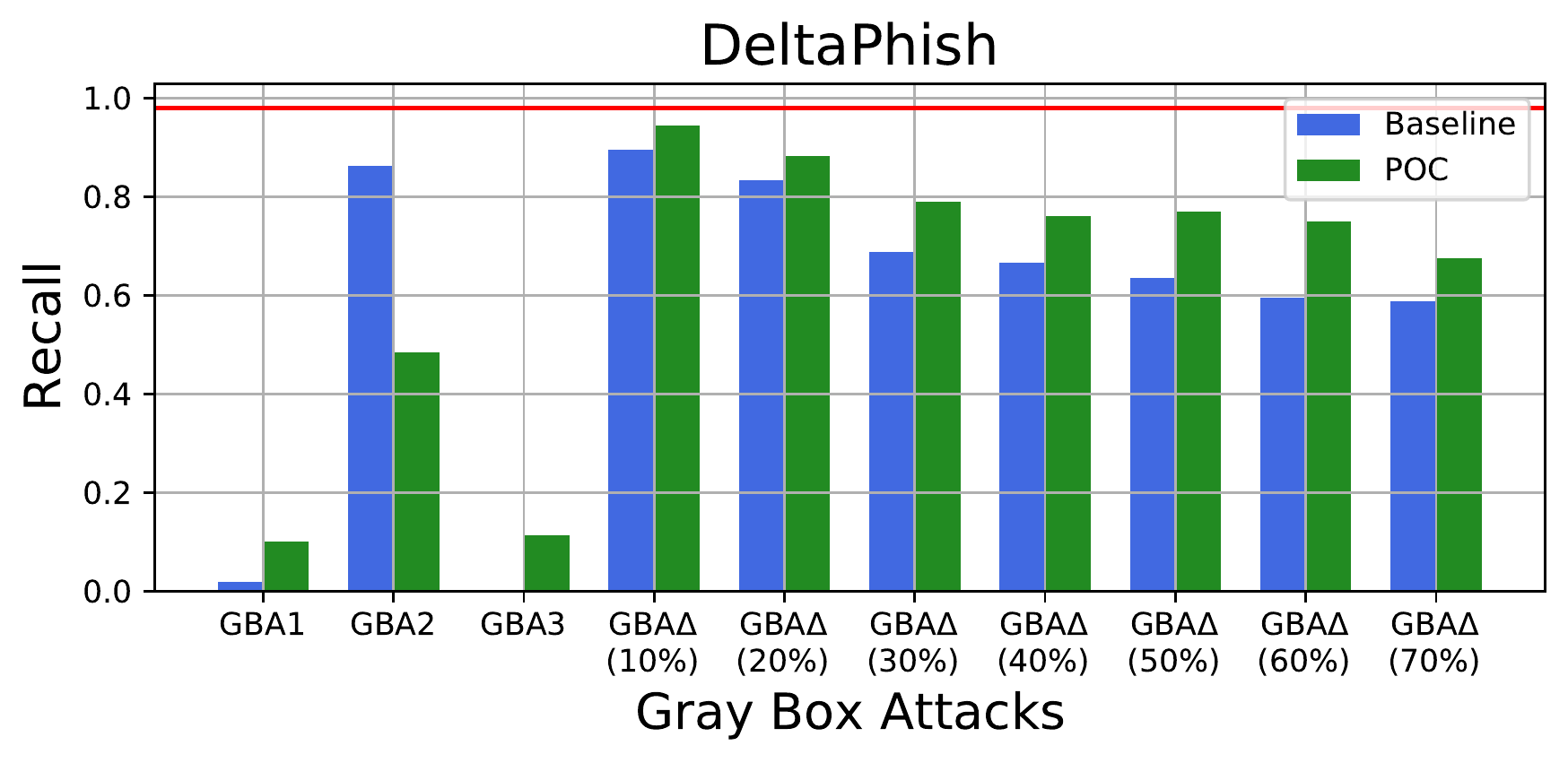}
        \caption{DeltaPhish. Best baseline: GB (F1-score: 0.99, and 0.98 for \POCscript).}        \label{sfig:pragmatic_deltaphish}
    \end{subfigure}\hfill%
    \begin{subfigure}{0.5\columnwidth}
        \centering
        \includegraphics[width=\linewidth]{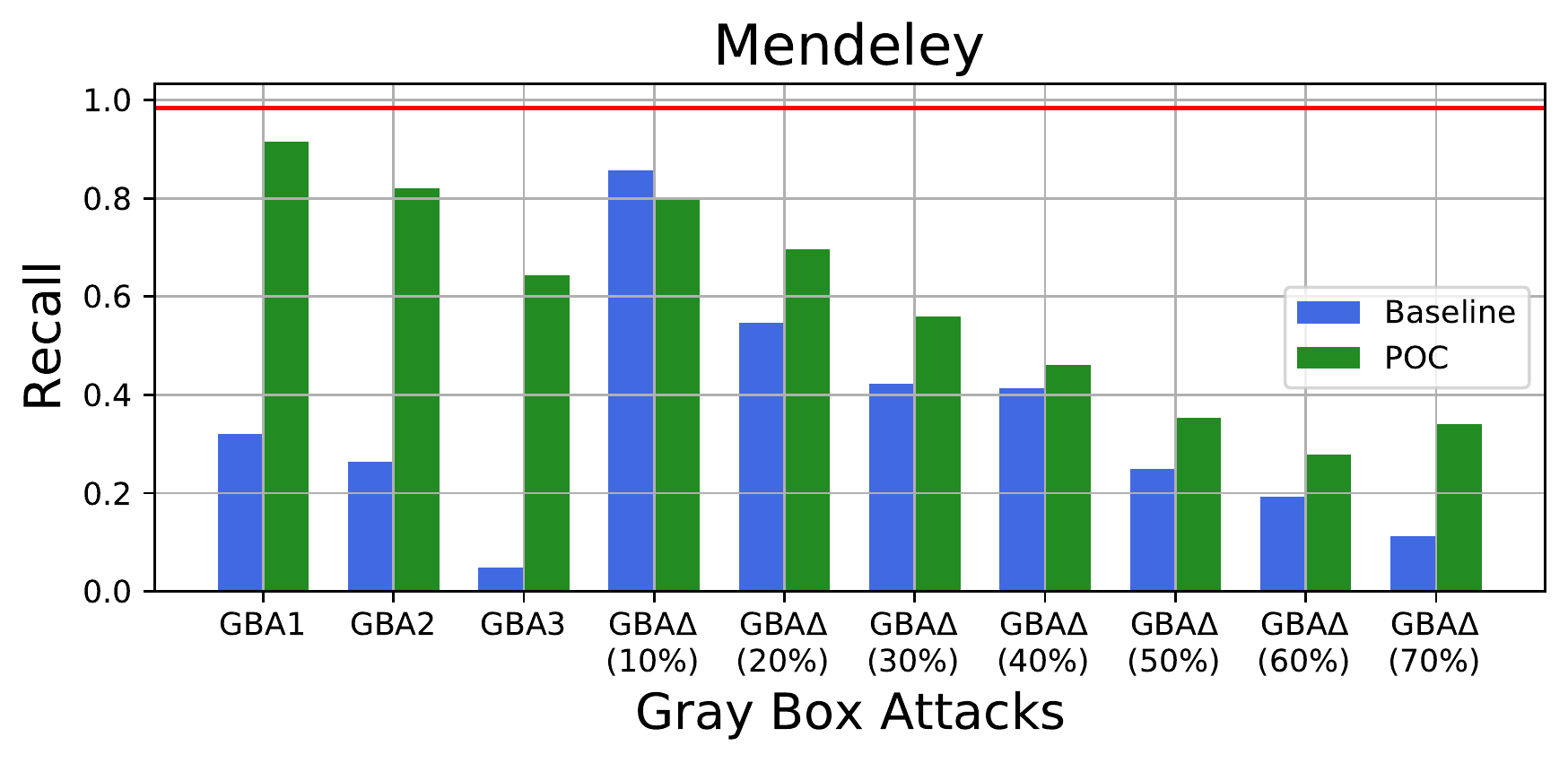}
        \caption{Mendeley. Best baseline: ET (F1-score: 0.99, and 0.96 for \POCscript).}
        \label{sfig:pragmatic_mendeley}
    \end{subfigure}\hfill%
        \begin{subfigure}{0.5\columnwidth}
        \centering
        \includegraphics[width=\linewidth]{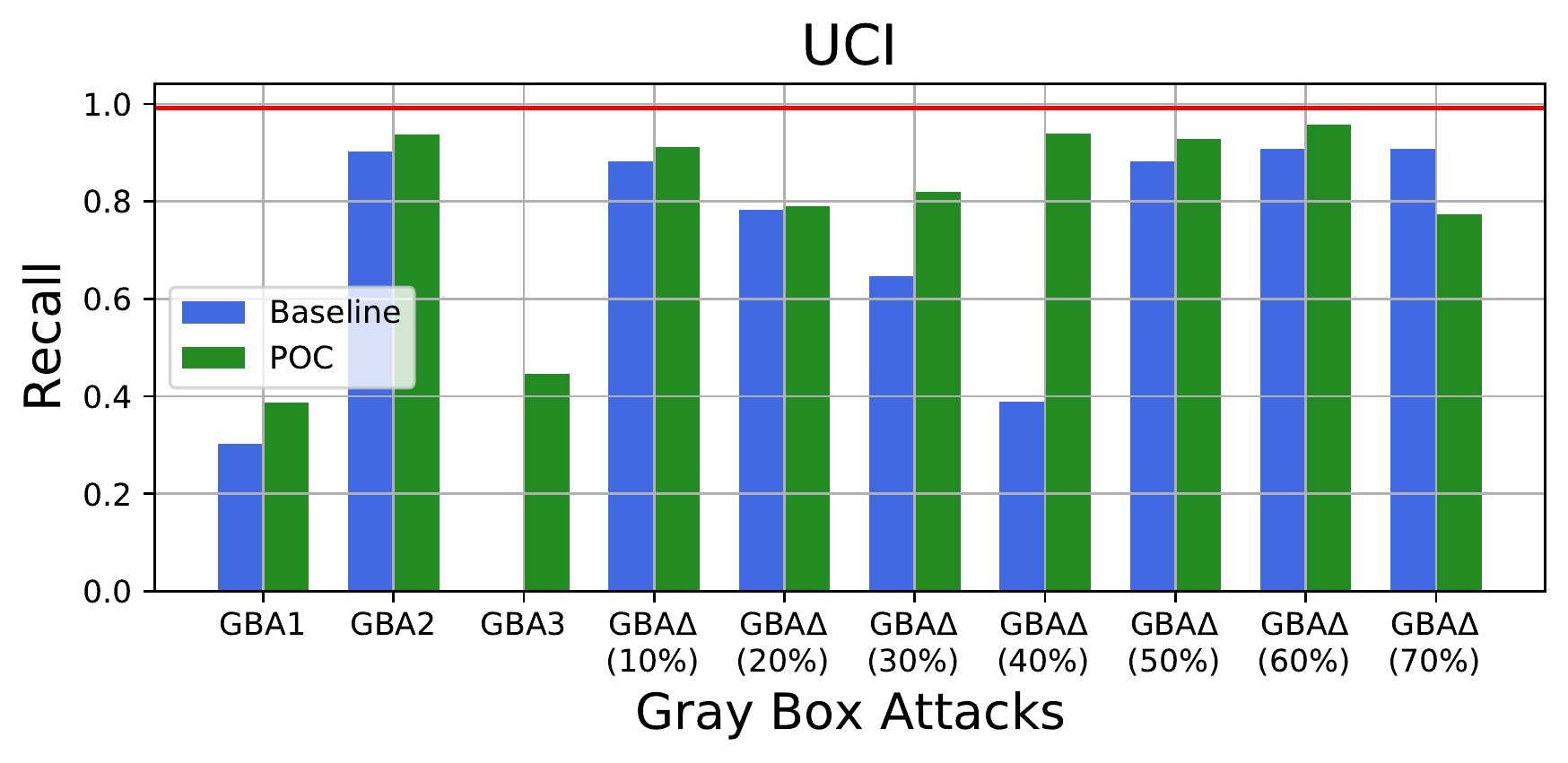}
        \caption{UCI. Best baseline: ET\\ (F1-score: 0.99, and 0.99 for \POCscript)}
        \label{sfig:pragmatic_uci}
    \end{subfigure}
    
    \caption{Effectiveness of attacks (as measured via the Recall) against the best baseline PD and its \POCscript-hardened version on each dataset. The red line on each subfigure is the Recall in the absence of attacks. The caption of each subfigure reports the F1-score in the no-attack case.}
    \label{fig:pragmatic}
\end{figure*}

We evaluated a huge number of classifiers in different conditions. However, in reality only a single classifier is used as PD---and the choice is made depending on its performance at training-time (i.e., in the no-attack scenario). Hence, we now investigate a pragmatic use case of \POC, where we analyze its benefits and tradeoffs when applied to `harden' the best classifier for each dataset (according to Table~\ref{tab:expt-baseline2}). 
Figure~\ref{fig:pragmatic} shows the Recall of the best baseline classifier alongside its \POC\ variant when they are subject to the Gray Box attacks considered in our paper. Each subfigure focuses on a specific dataset, and the red line in each subfigure reports the detection rate in the no-attack case. We analyze these subfigures, and then make some final recommendations.

\subsubsection{\LNU\ analysis}
\label{sssec:lnu_analysis}
Figure~\ref{sfig:pragmatic_lnu} focuses on the \LNU\ dataset, where the best classifier is GB which obtains near perfect performance---a result shared by its \POC\ variant. However, we can see that the latter is significantly more robust against \GBA{1} as well as against two \GBA{$\Delta$} (where $\Delta\!=\!60\%$ or $70\%$), as the Recall is above 10\% superior. 
Against all other attacks, the detection rate is either equivalent, or marginally superior than the baseline (up to 5\% increased Recall).

\subsubsection{DeltaPhish analysis} 
Figure~\ref{sfig:pragmatic_deltaphish} focuses on the DeltaPhish dataset, where the best baseline classifier is also GB, whose \POC\ variant has only a 0.01 less F1-score in the no-attack case. On this dataset, the performance of \POC\ is slightly inferior to the baseline against all the \GBA{$\Delta$} attacks, but significant differences arise in the simple attacks: for \GBA{1} and \GBA{3}, the baseline does not detect any attack whereas \POC\ can detect a small amount ($8\%$). In contrast, \GBA{2} barely affects the baseline GB but half of its samples can evade \POC. This is the most significant `defeat' of \POC---although its application on other strong baselines can be beneficial (e.g., for the deep learning MLP classifier, the Recall against \GBA{2} of \POC\ is 0.86 vs 0.77).

\subsubsection{Mendeley analysis} 
Figure~\ref{sfig:pragmatic_mendeley} focuses on the Mendeley dataset, where the best baseline classifier is ET.  \POC\ provides a significant mitigation (above 10\% better Recall) against 8 out of 10 attacks: the only exceptions are \GBA{$\Delta$} with $\Delta=20\%$, where the improvement is of smaller entity (4\%), and \GBA{2} where it is slightly worse than the baseline (by about 5\%). Specifically, \POC\ is barely affected by \GBA{1}, as it can successfully detect this attack with 0.85 Recall against the 0.37 of the baseline. All these benefits come at the `cost' of a 0.03 reduction in F1-score when no adversarial attack occurs.

\subsubsection{UCI analysis} 
Figure~\ref{sfig:pragmatic_uci} focuses on the UCI dataset, where the best baseline classifier is ET, whose hardened \POC\ variant achieves the same performance in the absence of attacks. 
We note that \POC\ exhibits weaker Recall (around 8\%) than the baseline only against \GBA{$\Delta$} with $\Delta=70\%$. \emph{In all other cases, \POC\ is superior.} Noteworthy are the successes against \GBA{$\Delta$} with $\Delta=40\%$, where \POC\ has a Recall of over 90\% against the 40\% of the baseline; and also against \GBA{3}, as the baseline cannot detect \textit{any} attack, whereas \POC\ can detect above 40\%.

\subsection{\POC\ Without Training}
\label{ssec:unrefined}
\begin{figure}[!htbp]
    \centering
    \begin{subfigure}{0.5\columnwidth}
        \centering
        \includegraphics[width=\linewidth]{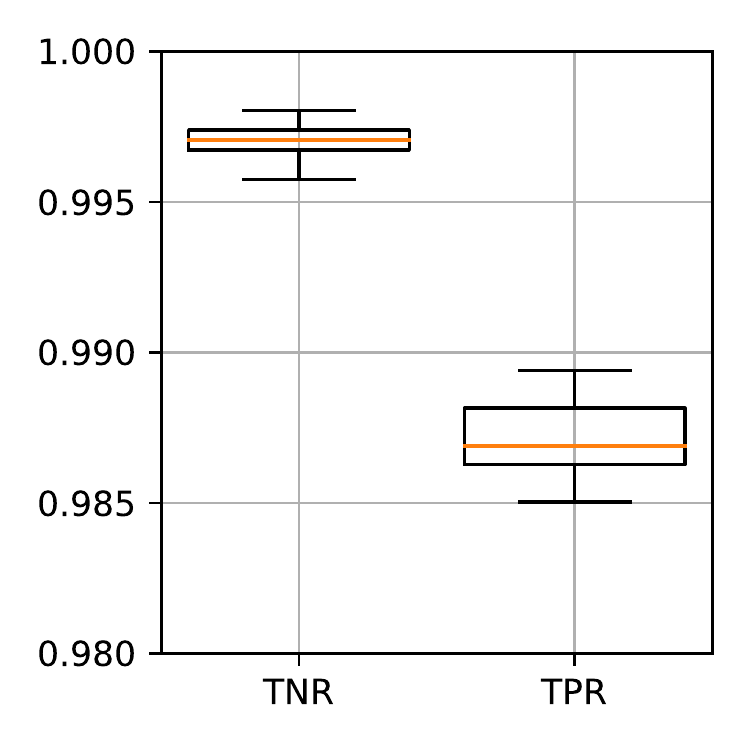}
        \caption{Results of 100 `blind' $\Psi$.}
        \label{sfig:tnr-tpr100}
    \end{subfigure}\hfill%
    \begin{subfigure}{0.5\columnwidth}
        \centering
        \includegraphics[width=\linewidth]{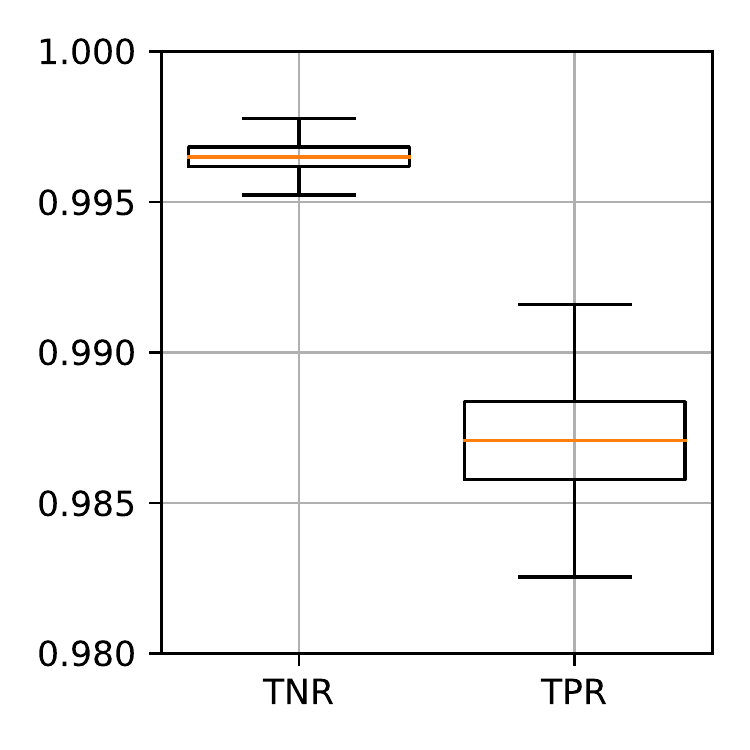}
        \caption{Results of 1000 `blind' $\Psi$.}        
        \label{sfig:tnr-tpr1000}
    \end{subfigure}
    \caption{Distribution of TNR and TPR achieved by `blindly' applying \POCscript\ to GB on \LNUscript\ (outliers are not shown).}
    \label{fig:tnr-tpr}
\end{figure}

We assess the performance of \POC\ when it is applied without training (i.e. without using an optimal choice of $\Psi$). To make the analysis humanly feasible, we focus on our proposed \LNU\ dataset, for which we consider the `best' baseline classifier: GB. We then ``blindly'' apply \POC\ 100 times to this baseline and then we re-do the experiment by applying \POC\ 1000 times. We perform this experiments with no training, and by using the same configuration parameters described in Section~\ref{sec:experiments} (i.e., having $\mathcal{P}=100\%$). The performance is measured by computing the TPR and TNR (i.e., 1-FPR) on the test-set (i.e., 20\% of \LNU). The results are shown in Figures~\ref{fig:tnr-tpr}.

The boxplots show that \POC\ yields practical performance even without training: the high TNR (always above 0.99) denotes low rates of false alarms, whereas the high TPR (always above 0.98) shows that phishing webpages are ably detected. It is encouraging that the distribution barely changes despite going from 100 to 1000 trials. \footnote{Of course, we do not claim this to be valid `anywhere-anytime', as such experiments focus on just a single configuration of a classifier (GB) on a single dataset (\LNUscript)}.

\subsection{Takeaway Message}
\label{ssec:takeaway}
By taking into account all above observations, we can draw the following conclusions. When used to harden the best classifier, \POC\ is a pragmatic solution against our Gray Box Attacks in 3 of 4 datasets. This is because \POC\ provides better (or same) adversarial robustness, but does not induce a significant performance drop in the absence of adversarial attacks. In contrast, on the DeltaPhish dataset, using \POC\ to harden the best baseline PD is not recommended: although it can detect some instances of \GBA{1} and \GBA{3} (against none of the baseline), the other results cannot justify its application to harden the ET classifier on this dataset. This is due to the specialized nature of the DeltaPhish dataset: while the other three datasets have malicious samples corresponding to `general' phishing webpages, the DeltaPhish dataset captures a \textit{specific} type of phishing attack. The entries in DeltaPhish are `legitimate' webpages that have fallen under the control of an attacker, which are different from  phishing  pages that are specifically created by an attacker. Hence, the high specificity of this dataset may yield a suboptimal hardening by \POC\ (on the very best baseline PD) against the proposed Gray Box Attacks.

Finally, we remark that \POC\ has---by definition---the additional benefit of yielding PDs that are hard to reverse engineer. This increases the difficulty of launching model stealing attacks, such as those conducted against the Google's Chrome phishing filter in~\cite{liang2016cracking}. Furthermore, even if an attacker were able to fully `crack' a \POC-hardened PD and infer its feature set, the attacker must repeat the process when the PD is periodically updated (to mitigate concept-drift~\cite{tian2018needle}) with more recent data, as the feature mapping is likely to change.

%% file: sections/8-prevalence.tex
\section{Experiments: \POC\ without complete prevalence}
\label{sec:prevalence}

As a final contribution of this paper, we evaluate the effectiveness of \POC\ when $\mathcal{P}(\mathbb{F},\Psi)\!<\!100\%$, meaning that some features of $\mathbb{F}$ are not included in any $oc$ composing $\Psi$. The expectation is that the robustness against attacks will increase (due to the `explicit' feature removal), but the performance in the absence of adversarial attacks will decrease because some information is lost (cf. Section~\ref{ssec:analysis}).

\subsection{Experimental Settings}
For simplicity, we perform experiments only on the \LNU\ dataset, where we consider the classifier yielding the best `baseline' PD---specifically, the GB classifier (cf. Section~\ref{sssec:lnu_analysis}).
The experimental settings are exactly the same as those described in Section~\ref{ssec:testbed}, but we do not require that $\mathcal{P}(\mathbb{F},\Psi)\!=\!100\%$. 
In particular, we assess \POC\ for different $\mathcal{P}$. Hence, we apply \POC\ so that the resulting $\mathcal{P}(\mathbb{F},\Psi)$ falls within 6 values ranging from 65\% to 90\% (at 5\% increments). As an example, since the \LNU\ dataset contains 27 features (cf. Table~\ref{tab:LNUfeatures}), when $\mathcal{P}(\mathbb{F},\Psi)\!=\!70\%$ it means that the \POC-hardened PD uses 19 features (across all its $oc$).

\subsection{Results}
\begin{figure}[!htbp]
    \centering
    \begin{subfigure}{0.5\columnwidth}
        \centering
        \includegraphics[width=\linewidth]{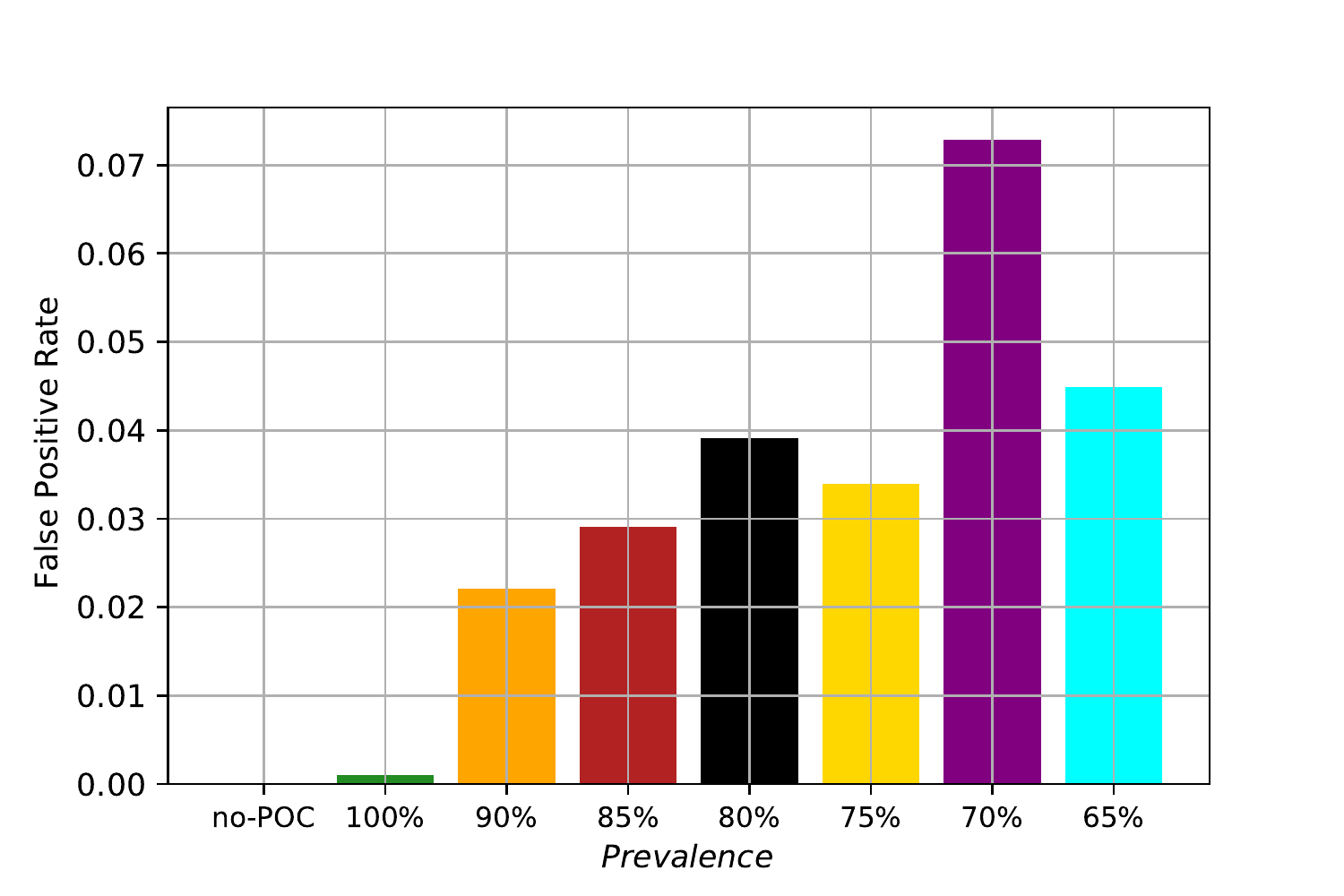}
        \caption{False Positive Rate.}
        \label{sfig:prevalence_fpr}
    \end{subfigure}\hfill%
    \begin{subfigure}{0.5\columnwidth}
        \centering
        \includegraphics[width=\linewidth]{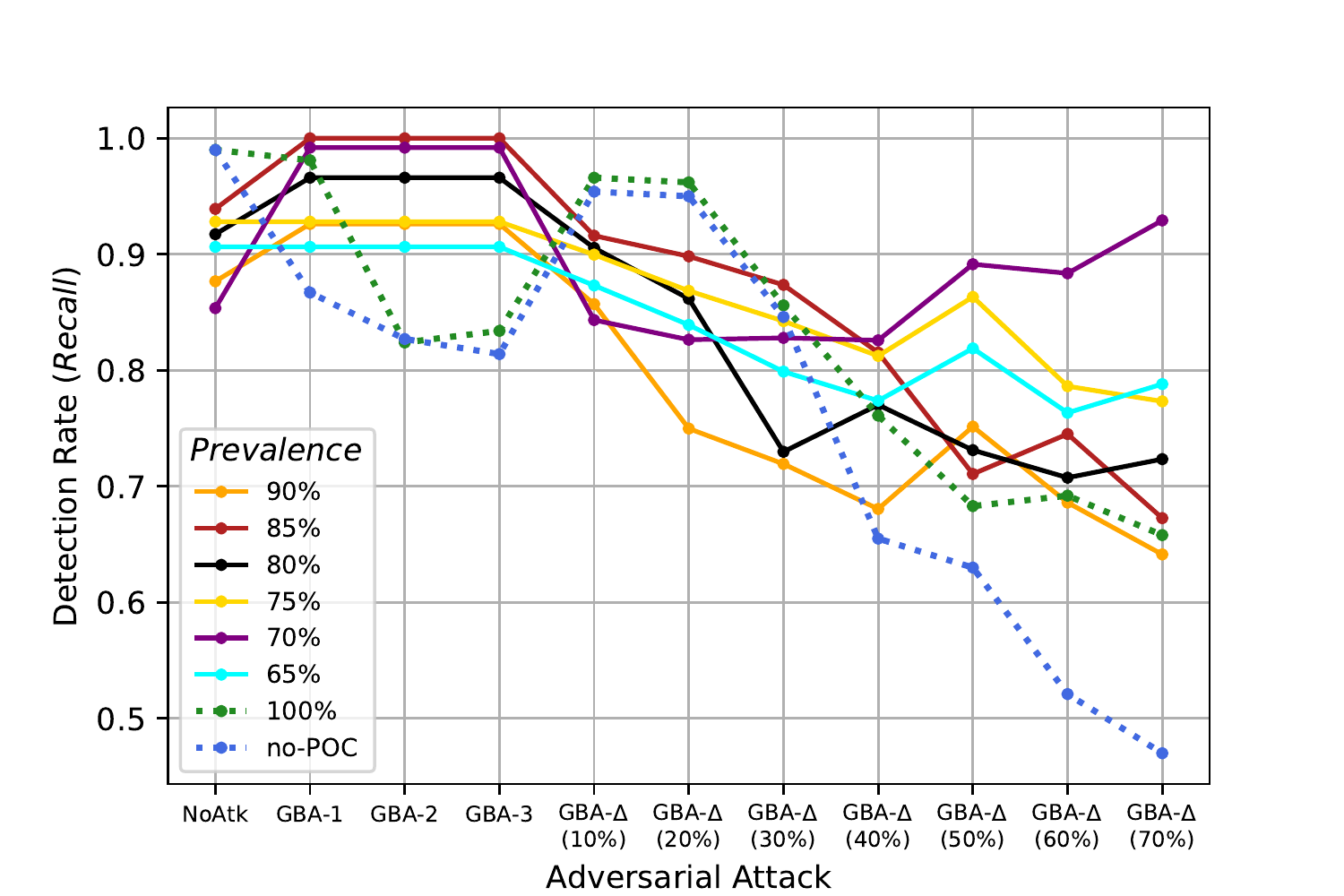}
        \caption{Detection Rate.}
        \label{sfig:prevalence_dr}
    \end{subfigure}\hfill%
    \caption{Performance of \POCscript\ with varying \textit{prevalence}.}
    \label{fig:prevalence}
\end{figure}

We evaluate all such \POC-hardened variants of the GB classifier both in the absence of attacks and against all the 10 adversarial attacks considered in our paper, and report the results in Figures~\ref{fig:prevalence}.

Figure~\ref{sfig:prevalence_fpr} shows the \textit{false positive rate} as a function of $\mathcal{P}$; where the two leftmost bars report the FPR of the `baseline' GB and the \POC-hardened GB with complete \textit{prevalence} (from Section~\ref{sec:experiments}). In contrast, Figure~\ref{sfig:prevalence_dr} shows the \textit{detection rate} against all the considered adversarial attacks (on the horizontal axis), as well as in the no-attack case (the leftmost value); the dotted lines represent the results reported in Section~\ref{sec:experiments} (included for comparison), whereas full lines represent the \POC-hardened PDs with varying \textit{prevalence}.

From Figure~\ref{fig:prevalence} we can see that---in the absence of attacks---the performance of \POC\ when $\mathcal{P}\!\leq\!90\%$ is worse with respect to the results shown in Section~\ref{sec:no_attack}. Indeed, when no attacks occur, the FPR (Figure~\ref{sfig:prevalence_fpr}) is higher and the detection rate (leftmost value in Figure~\ref{sfig:prevalence_dr}) is also inferior. This is due to the loss of information induced when $\mathcal{P}\!<\!100\%$.
However, such increased FPR is ably compensated by the greater detection rate in the presence of adversarial attacks. As shown in Figure~\ref{sfig:prevalence_dr}. with the sole exception of \GBA{$\Delta$} where $\Delta\!$=10 or 20\%, the full lines denote better results than the dotted lines. As an example, when $\mathcal{P}\!=\!70\%$, the corresponding PD is not affected at all by the simple attacks, and its detection rate never goes below 83\%---but, it also achieves the greatest FPR (0.073 according to Figure~\ref{sfig:prevalence_fpr}). 

In summary, these results match our expectations. From a practical perspective, using \POC\ without complete \textit{prevalence} is beneficial if a PD is likely to be targeted by the proposed Gray Box attacks, and if the deployment setting can accept a slightly worse performance in the absence of such attacks.

%% file: sections/9-conclusions.tex

\section{Conclusions}
\label{sec:conclusions}
It is clear from ProofPoint's 2020 ``State of the Phish'' report that despite decades of work to counter phishing attacks, phishing represents a major attack vector for malicious hackers. 

In this paper, we propose a series of complex and simple Gray Box attacks on existing machine learning based classifiers for phishing website detection. We formally define the Impact of an attack on a dataset and classifier in terms of the percentage drop in predictive performance and show that these attacks cause a significant drop in performance of past work using ML classifiers. 

We develop the \POC\ algorithm that uses a mix of randomization (to reduce the probability that the adversary can guess the features used) and feature transformation (to further reduce this probability). We show that \POC---despite not representing a universal panacea against adversarial attacks---is more robust against all the considered Gray Box attacks than past classifiers, and does not degrade their performance in the absence of such attacks.

Our paper considers 13 classifiers (including new classifiers such as Google's Deep \&\ Wide that have not been used previously for phishing detectors to the best of our knowledge) using 4 datasets (including the new \LNU\ dataset that we release as an additional contribution of this paper). In contrast, most past work on adversarial phishing detectors consider only one dataset and one classifier.
\\ \\
\noindent\textbf{Acknowledgements.} We thank the anonymous referees for their excellent comments. We are also grateful to ONR grant N00014-20-1-2407.

%% file: main.bbl

%% file: biographies.tex
\newpage
\begin{IEEEbiography}
  [{\includegraphics[width=1in,height=1.25in,keepaspectratio]{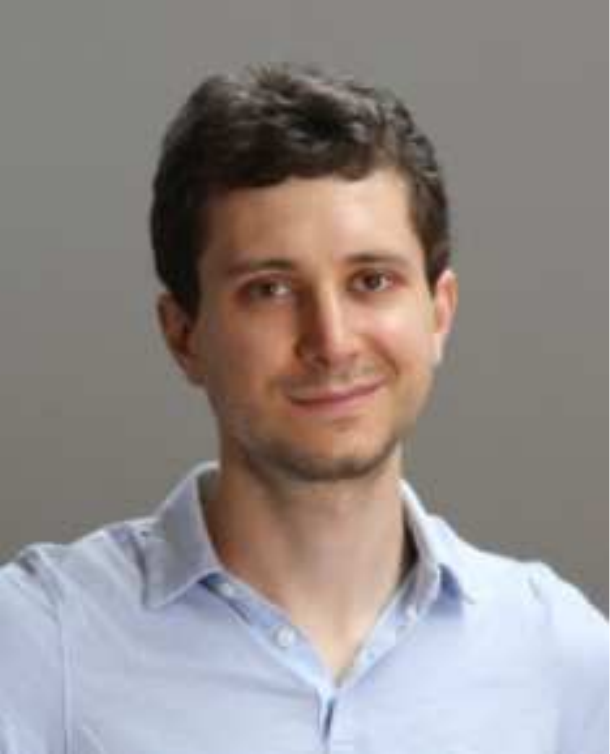}}]{Giovanni Apruzzese} is an Assistant Professor within the Hilti Chair of Data and Application Security at the University of Liechtenstein since 2022, and was previously a PostDoc at the same institution since 2020. He received the PhD Degree and the Master's Degree in Computer Engineering (summa cum laude) in 2020 and 2016 respectively at the Department of Engineering ``Enzo Ferrari'', University of Modena and Reggio Emilia, Italy. In 2019 he spent 6 months as a Visiting Researcher at Dartmouth College (Hanover, NH, USA) under the supervision of Prof. V.S. Subrahmanian. His research interests involve all aspects of big data security analytics with a focus on machine learning, and his main expertise lies in the analysis of Network Intrusions, Phishing, and Adversarial Attacks.
  
  Homepage: \url{https://giovanniapruzzzese.com}
\end{IEEEbiography}
\vspace{-5em}
\begin{IEEEbiography}
 [{\includegraphics[width=1in,height=1.25in,clip,keepaspectratio]{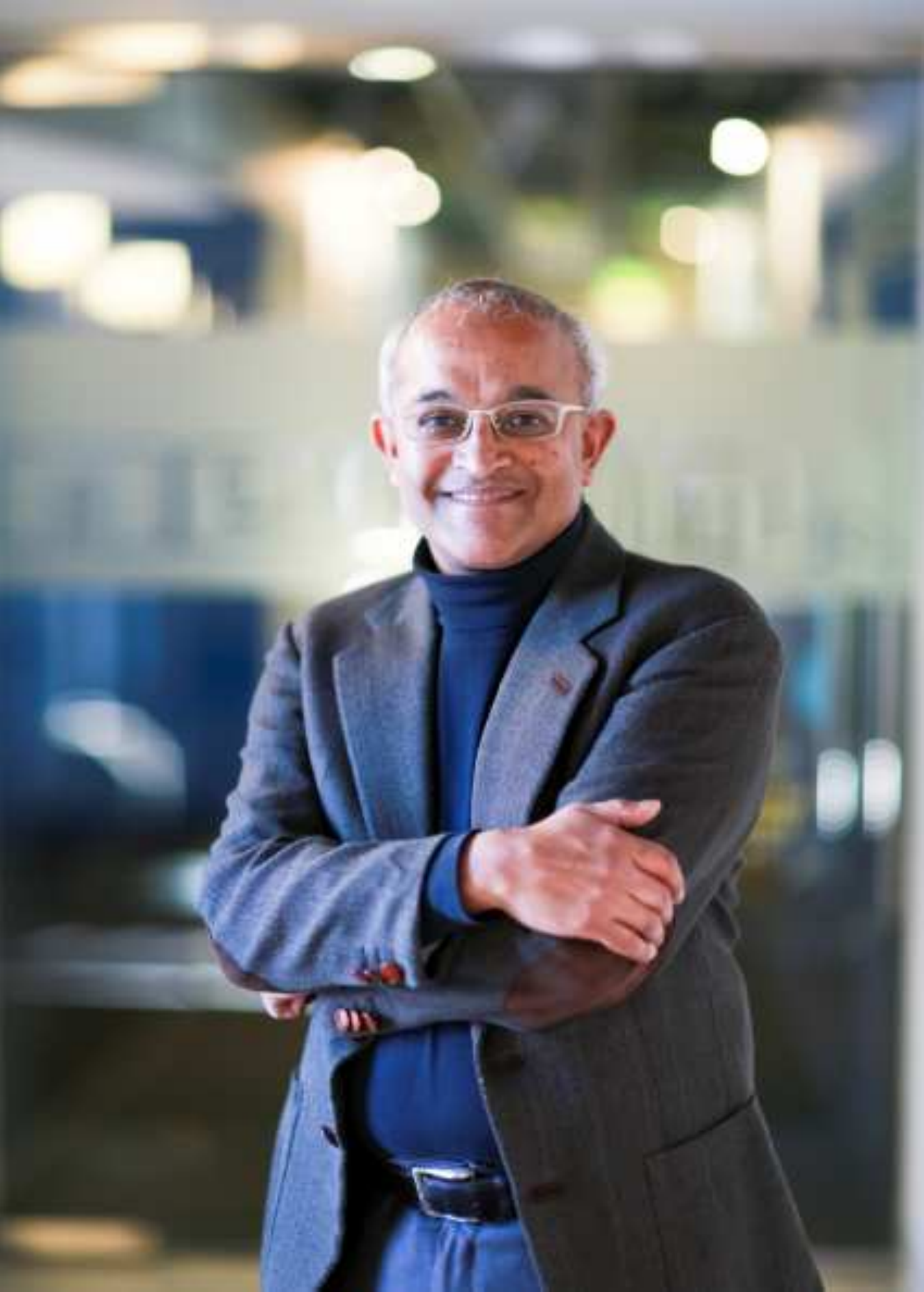}}]{V.S. Subrahmanian} is the Walter P. Murphy Professor of Computer Science and Buffett Faculty Fellow in the Buffett Institute of Global Affairs at Northwestern University. He was previously the Dartmouth College Distinguished Professor in Cybersecurity, Technology, and Society and Director of the Institute for Security, Technology, and Society at Dartmouth. Before that, he was a Professor of Computer Science at the University of Maryland from 1989-2017 where he also served for 6+ years as Director of the University of Maryland’s Institute for Advanced Computer Studies. Prof. Subrahmanian is an expert on big data analytics including methods to analyze text/geospatial/relational/social network data, learn behavioral models from the data, forecast actions, and influence behaviors with applications to cybersecurity and counterterrorism. He has written five books, edited ten, and published over 300 refereed articles. He is a Fellow of the American Association for the Advancement of Science and the Association for the Advancement of Artificial Intelligence and received numerous other honors and awards. His work has been featured in numerous outlets such as the Baltimore Sun, the Economist, Science, Nature, the Washington Post, American Public Media. He serves on the editorial boards of numerous journals including Science, the Board of Directors of the Development Gateway Foundation (set up by the World Bank), SentiMetrix, Inc., and on the Research Advisory Board of Tata Consultancy Services. He previously served on DARPA’s Executive Advisory Council on Advanced Logistics and as an ad-hoc member of the US Air Force Science Advisory Board. 
 
 Homepage: \url{https://vssubrah.github.io/}
\end{IEEEbiography}